\documentclass{article}
\usepackage{amsmath,amssymb,amsfonts}

\numberwithin{equation}{section}

\newcommand{\phm}[1]{\phantom{#1}}

\def\al{\alpha}
\def\be{\beta}
\def\de{\delta}
\def\ep{\epsilon}
\def\la{\lambda}
\def\La{\Lambda}
\def\si{\sigma}
\def\ga{\gamma}
\def\Lama{\Lambda_{(\al)}}
\def\Ja{J_{(\al)}}
\def\j{J_{(1)}}
\def\jj{J_{(2)}}
\def\tha{\theta_{(\al)}}
\def\th{\theta_{(1)}}
\def\Th{\theta_{(2)}}
\def\pha{\phi_{(\al)}}
\def\Ia{I_{(\al)}}
\def\eps{\varepsilon}
\def\om{\omega_{(1)}}
\def\Om{\omega_{(2)}}
\def\oma{\omega_{(\al)}}
\def\etaa{\eta_{(\al)}}
\def\mphi{\varphi}
\def\mtheta{\vartheta}
\def\z{z_{(1)}}
\def\Z{z_{(2)}}
\def\za{z_{(\al)}}
\def\Am{\mathsf{A}}
\def\bi{\bar{\imath}}
\def\ti{\hat{\imath}}
\def\ic{\mbox{i}}
\def\si{\sigma}
\def\H{{\cal H}}
\def\F{{\cal F}}
\def\C{{\cal C}}
\def\G{{\cal G}}
\def\J{\mathsf{J}}
\def\O{\mathsf{O}}
\def\DD{\mathsf{D}}
\def\x{\mathbf{x}}
\def\etazp{\eta_{\text{ZP}}}
\def\etac{\eta_{\text{C}}}
\def\dif{\, \mathrm{d}}
\def\zetaop{\zeta_0^\prime}
\def\rhoop{\rho_0^\prime}
\def\Gd{\mathsf{G}}
\def\Gi{\tilde{\mathsf{G}}}
\def\epsi{\eps_I}
\def\epsr{\eps_R}
\def\Pr{{\cal P}}
\def\tr{^\mathrm{T}}
\newcommand{\dfun}[2]{\frac{\de #1}{\de #2}}
\newcommand{\dpar}[2]{\frac{\partial #1}{\partial #2}}
\newcommand{\dt}[2]{\frac{\mathrm{d} #1}{\mathrm{d} #2}}
\newcommand{\D}[2]{D_{\mbox{ }#1}^{#2}}
\newcommand{\stc}[2]{c_{\mbox{  }#2}^{#1}\,}
\newcommand{\hD}[2]{\hat{D}_{\mbox{ }#1}^{#2}}
\newcommand{\xiv}[2]{\chi^{#1}_{#2}}
\newcommand{\xif}[2]{\chi_{#1}^{#2}}
\newcommand{\A}[3]{A_{#1#2}^{#3}\,}

\begin{document}

\title{\textbf{Weakly nonlinear dynamics in noncanonical Hamiltonian
systems with applications to fluids and plasmas}} 
\author{P.\ J.\ Morrison$^\dagger$ and J.\ Vanneste$^\ddagger$}
\date{{\normalsize $^\dagger$ Dept.\ of Physics and Institute for Fusion Studies,
University of Texas, Austin, USA \\
      $^\ddagger$ School of Mathematics and Maxwell Institute for Mathematical Sciences, University of Edinburgh, UK}}

\maketitle

\begin{abstract}
A method, called {\it beatification}, is presented for rapidly extracting weakly nonlinear Hamiltonian systems that describe the dynamics near equilibria for systems  possessing Hamiltonian form in terms of noncanonical Poisson brackets.  The procedure applies to systems like fluids and plasmas in terms of Eulerian variables that have such noncanonical Poisson brackets, i.e., brackets with nonstandard and  possibly degenerate  form.  A  collection of examples of both finite and infinite dimensions is presented. 

\end{abstract}



\section{Introduction}

The most important physical systems are either governed by evolution equations that are, in one sense or another, Hamiltonian or possess Hamiltonian limits in which dissipative or transport terms associated with phenomenological constants are dropped.  This imposes strong constraints on their dynamics with, for instance the existence of differential invariants, and the possibility of symmetry-related conservation laws.  The existence of a Hamiltonian structure is particularly important for perturbative problems. The structure can not only be exploited to simplify asymptotic developments considerably; it also leads to several results, such as the adiabatic invariance of action, without counterparts in non-Hamiltonian systems. 

Hamiltonian perturbation has been largely developed for canonical systems, with equations
\[ 
\dot{p}_{a} = \dpar{H}{q^{a}}= \{p_a,H\}, \quad \dot{q}^{a}=-\dpar{H}{p_{a}}= \{q^a,H\}, 
\]
associated with the Hamiltonian $H$ and canonical Poisson bracket 
\begin{equation}
\{f,g\}=\dpar{f}{p_{a}} \dpar{g}{q^{a}}-\dpar{g}{p_{a}} \dpar{f}{q^{a}}\,,
\label{cPB}
\end{equation}
where $a=1\dots,N$ and repeated indices here and henceforth  are summed. The perturbative methods typically rely on canonical transformation, which, after truncation, simplify the Hamiltonian while leaving the form of the Poisson bracket unchanged. This is particularly efficient since the computation is focussed on a scalar --- the Hamiltonian --- rather than on the evolution equations themselves. The approach is well illustrated by the classical results on averaging, passage through resonance, adiabatic invariance, etc. (e.g.\ \cite{AKN88,hen93}).

Here, we are interested in the more general class of noncanonical Hamiltonian systems (a terminology introduced  \cite{MG80}, see also e.g.\ \cite{mor82,mor98}) that have the form 
\begin{equation} \label{eqn:hamilton}
\dot{z}^i = J^{ij}(z) \dpar{H}{z^j}
=\{{z}^i,H\},
\end{equation}
where $H$ is the Hamiltonian,   $i,j=1,2,\ldots,M$, and  
 $J^{ij}$ denotes the components of the  Poisson matrix, which is  skew-symmetric, satisfies the Jacobi identity
\begin{equation} \label{eqn:jacobi}
J^{ik} \dpar{J^{jl}}{z^k} +J^{jk} \dpar{J^{li}}{z^k} +  J^{lk} \dpar{J^{ij}}{z^k} = 0,
\end{equation}
and defines the Poisson bracket
\begin{equation}
\{f,g\} = \dpar{f}{z^i} J^{ij} \dpar{g}{z^j}.
\label{PB}
\end{equation}
In general, $J^{ij}$ depends on $z$ and is often degenerate (singular), unlike the canonical form where it  is given by
\begin{equation}
J_c = \left(
\begin{array}{rl}
0_N & I_N \\ 
-I_N & 0_N
\end{array}
\right)
\,, 
 \label{canJ}
 \end{equation} 
with  $I_N$    the $N\times N$ identity matrix and $0_N$ an   $N\times N$ block of zeros.  Substitution of (\ref{canJ}) into (\ref{PB}) yields (\ref{cPB}).

For the degenerate case, it   was known by S.\ Lie  (see e.g.\ \cite{car,eis}) that the null-space of $J$ is spanned by the gradients of the Casimir functionals $C^{\al}$ (referred to as distinguished functions by Lie) that satisfy
\begin{equation} \label{eqn:casimir}
J^{ij} \dpar{C^{\al}}{z^j}=0, \quad \alpha=1,2,\ldots,M-2N\,,
\end{equation} 
where $2N$ is the rank of $J^{ij}$.   Note,  we will systematically use greek symbols to index the Casimirs functionals in what follows.   Systems of the form (\ref{eqn:hamilton}) emerge in many areas of physics, such as  fluid and plasma models  in terms of Eulerian variables.   They are  typically obtained from canonical systems after symmetry reduction, and have therefore the advantage of a state space of reduced dimensionality  \cite{mor82,mor98,MW74,MR}.

However, when in comes to the development of perturbation theories, the form (\ref{eqn:hamilton}) is generally considered inconvenient, notwithstanding the efficacious  use of noncanonical coordinates for perturbation theory in some cases \cite{lit79}. The difficulty stems from the $z$-dependence of $J^{ij}(z)$: when $z$ is expanded, for instance in power series of a small parameter,  $J^{ij}(z)$ also needs to be expanded. If this expansion is then truncated, the Jacobi identity (\ref{eqn:jacobi}) then usually ceases to hold, and the system loses its Hamiltonian nature. 

In this paper, we discuss how this difficulty can be overcome in order to study perturbatively the weakly nonlinear dynamics of noncanonical systems in the neighbourhood of a (stable) equilibrium $z_{0}$. The idea is to perform a near identity change of variables that  transforms $J^{ij}(z)$ into its (constant) value $J^{ij}(z_{0})$ at the equilibrium. Because transforming the Poisson matrix to a constant, i.e., a form independent of $z$,  is the first step of canonization, transforming to canonical form, we refer to this procedure as {\it beatification}.  This change of variables is then introduced in the Hamiltonian, which can be truncated. The result is a Hamiltonian system with constant (but noncanonical) Poisson matrix, where nonlinearity has been transferred from the Poisson bracket to the Hamiltonian. This system can then be studied perturbatively, by applying further changes of variables that either leave the constant Poisson matrix invariant or transform it into the canonical form of (\ref{canJ}), which is  
known to be always possible by Darboux's theorem (e.g.\ \cite{sam95}). Here, we provide a practical implementation for systems near an equilibrium. 
In principle, the approach can be carried out order by order in the small parameter that measures the size of the perturbation away from the equilibrium. We limit our discussion to the first step in the procedure. This is sufficient to describe the leading-order effect of the nonlinearity and study effects such as quadratic resonance and possible explosive instabilities, a special case of which was considered in detail in \cite{KM95}. 

We note, the linearization problem of  literature  \cite{wei83,con85,AG90}  of smoothly transforming general brackets to local Lie-Poisson form, differs from the procedure here: our goal is to `flatten' brackets to  first order and  thereby remove the linear dependence replacing it by a constant. 

The paper is organized as follows.  In Section \ref{sec:finite} we consider  finite-dimensional systems of the form (\ref{eqn:hamilton}). For these,   an explicit formula is given for the change of variable that makes the Poisson matrix  constant to leading order. We consider both the cases of nondegenerate and degenerate Poisson matrices, and we give a simplified formulation for systems  with  Lie-Poisson structure. Applications based on the real semi-simple Lie algebras $\mathfrak{so}(2,1)$ and $\mathfrak{so}(3)$,  where the former applies to the Kida vortex and the latter to  the rigid body,  the heavy  top, and a two-spin system, are presented in Section \ref{sec:app} to illustrate the method.  Infinite-dimensional systems are discussed in Section \ref{sec:infinite}. For these, we do not give a general form for the change of variables,  which would be overly complicated. Rather, we derive the transformation explicitly for  Poisson brackets  of  fluid dynamics and plasma physics  when the equilibrium depends on only a single coordinate.  In  Section \ref{sec:eul} we first  consider  the Lie-Poisson  bracket  for the  two-dimensional  Euler equation that describes  the ideal  fluid,  which is identical to the bracket that describes the one-dimensional Vlasov--Poisson and other systems. Then, in  Section \ref{sec:kir},  we consider the  bracket  for two-dimensional stratified fluids and for two-dimensional  magnetohydrodynamics. For these four physical systems, we implement the relevant change of variables in the Hamiltonian and derive the equations of motion in the transformed variables. These equations are suited for the study of the weakly nonlinear interactions of waves or more generally modes, for instance in the statistical treatment of weak turbulence (e.g.\ \cite{davidson,zakharov}).  We conclude in Section \ref{sec:discussion} where we summarize and discuss our results and consider some future applications.  The paper also contains several appendices that expand upon material presented in the main text.  In particular,  Appendix \ref{sec:canon} demonstrates how a canonical structure for the weakly nonlinear Vlasov-Poisson equation can be obtained by building on the coordinate transformation of Section  \ref{sec:kir}  and on earlier work on the canonization of the linearized system.


\section{Finite-dimensional systems} \label{sec:finite}

Our treatment of finite-dimensional systems is divided into three subsections.  In Section \ref{ssec:formulation} we set up the basic expansion formulation, in preparation for Section \ref{ssec:flat} where beatification is explicitly undertaken.  Then, in Section \ref{ssec:lpsystems} we consider the special case of Lie-Poisson systems where the initial Poisson vector is linear in $z$.


\subsection{Formulation}
\label{ssec:formulation}

An equilibrium $z_0$ of (\ref{eqn:hamilton}) satisfies 
\begin{equation}
\dot{z}^i =J^{ij}_0 \dpar{H_0}{z_0^j} := J^{ij}(z_0) \dpar{H(z_0)}{z_0^j} = 0.
\end{equation}
Provided that the rank of $J^{ij}$ does not change at $z_0$, $z_0$ is a critical point of the combination
\[
F(z) := H(z) + \la_{\al}(z_0) \, C^{\al}(z)
\]
for some $\la_{\al}$, i.e.
\[
\dpar{F(z_0)}{z_0^j}=0.
\]
To study the evolution of a small disturbance to $z_0$, we write
\begin{equation}
z = z_0 + \ep z^\prime,
\end{equation}
where $\ep \ll 1$. In what follows, we will omit the prime, using $z$ for the disturbance. Its evolution equation is governed by
\[
\dot{z}^i = J_z^{ij} \dpar{F_z}{z^j},
\]
where
\begin{equation} \label{eqn:freeen}
J_z^{ij} := J^{ij}(z_0 + \ep z)\quad \mbox{ and } \quad F_z := \ep^{-2} \left[
F(z_0 + \ep z)-F(z_0)\right].
\end{equation}
The Hamiltonian $F_z$, which we will refer to as the free energy  is
sometimes called the pseudoenergy. 
The $O(\ep)$ approximation of this equation reads
\begin{equation} \label{eqn:dist}
\dot{z}^i = \left( J_0^{ij} + \ep \dpar{J_0^{ij}}{z_0^l} z^l \right)
	\left( \dpar{^2 F_0}{z_0^j \partial z_0^k} z^k
		+ \frac{\ep}{2} \dpar{^3 F_0}{z_0^j \partial z_0^k \partial z_0^m} z^k z^m \right) + O(\ep^2).
\end{equation}

In general, this equation is not Hamiltonian: due to the truncation of $J^{ij}_z$ as
\begin{equation} \label{eqn:jz}
J_0^{ij} + \ep \dpar{J_0^{ij}}{z_0^l} z^l,
\end{equation}
the Jacobi identity is not satisfied. However, through a change of variable, it is possible to find a new equation, equivalent to (\ref{eqn:dist}) up
to $O(\ep^2)$, that  is Hamiltonian. Let $\eta = \eta(z)$ be the new dependent variable. The corresponding Poisson matrix   is given by
\begin{equation} \label{eqn:jeta}
J^{ij}_\eta = \dpar{\eta^i}{z^k} J_z^{kl} \dpar{\eta^j}{z^l}.
\end{equation}
We will choose $\eta$ such that
\begin{equation} \label{eqn:cond}
J_\eta^{ij} = J_0^{ij} + O(\ep^2).
\end{equation}
In contrast to  the truncation of $J^{ij}_z$  given by (\ref{eqn:jz}), the truncation of $J^{ij}_\eta$ to $O(\ep)$ satisfies the Jacobi identity: indeed, it is simply given by $J^{ij}_0$ and thus has constant coefficients, making satisfaction of  (\ref{eqn:jacobi}) immediate. The evolution equation for $\eta$ takes the Hamiltonian form
\begin{equation} \label{eqn:eta}
\dot{\eta}^i = J_0^{ij} \dpar{F_\eta}{\eta^j} + O(\ep^2),
\end{equation}
where the Hamiltonian
\[
F_\eta := \ep^{-2} \left[ F(z_0 + \ep z(\eta))-F(z_0)\right]
\]
can be consistently truncated to $O(\ep)$.


\subsection{Beatification:  flattening the Poisson matrix}
\label{ssec:flat}

We now construct the new variable $\eta$ leading to (\ref{eqn:cond}). The change of variable is clearly near-identity so that, to the order of interest, it can be written
\begin{equation} \label{eqn:cv}
\eta^i = z^i + \frac{\ep}{2} \, \D{kl}{i} \, z^kz^l + O(\ep^2),
\end{equation}
where the quantity  $\D{kl}{i}$ is symmetric in $kl$. Introducing this expression in (\ref{eqn:jeta}), we can rewrite the condition (\ref{eqn:cond}) as
\begin{equation} \label{eqn:condd}
S^{ij}_{\phm i \phm j l}:=J_0^{ik} \D{kl}{j} + J_0^{kj} \D{kl}{i} + \dpar{J_0^{ij}}{z_0^l}=0.
\end{equation}
Because of the skew-symmetry of $J^{ij}_0$, this system contains a maximum of
$M^2 (M-1)/2$ independent equations, while $\D{kl}{i}$ contains $M^2 (M+1)/2$
coefficients, and their  difference,  $M^2$, arises because of  the nonuniqueness of the variables we seek.  Equation (\ref{eqn:condd}) is akin to the formula obtained in Riemannian geometry when one seeks coordinates, so-called normal coordinates,  in which the metric is flat (e.g.\ \cite{bir}).  In this case the quantities analogous to $\D{kl}{i}$ turn out to be the Christoffel symbols.  In our case, we will see that the situation is somewhat more complicated.

To solve (\ref{eqn:condd}) for $\D{kl}{i}$, we introduce the following decomposition: 
\begin{equation} \label{eqn:decomp}
\D{kl}{i} = \frac1{3}\dpar{J_0^{im}}{z_0^k} \omega_{lm} + \frac1{3}\dpar{J_0^{im}}{z_0^l} \omega_{km}
		+ \hD{kl}{i},
\end{equation}
where $\omega_{lm}$ and $\hD{kl}{i}$ are yet to be determined; 
$\omega_{lm}$ is assumed to be skew-symmetric, while $\hD{kl}{i}$ is symmetric in $kl$. Introducing  expression (\ref{eqn:decomp}) into (\ref{eqn:condd}) yields
\begin{eqnarray} \label{eqn:compl}
S^{ij}_{\ \ l}&=&
\frac{1}{3} \dpar{J_0^{ij}}{z_0^k}  \left(\de^k_{\phm k l} - J_0^{km} \omega_{ml}\right) 
+  \frac{1}{3} \dpar{J_0^{im}}{z_0^l}  \left(\de^j_{\phm j m} - J_0^{jk} \omega_{km}\right) 
+  \frac{1}{3} \dpar{J_0^{mj}}{z_0^l}  \left(\de^i_{\phm i m} -J_0^{ik} \omega_{km} \right) 
  \nonumber\\
 &{\ }& \hspace{4cm} +J_0^{ik} \hD{kl}{j} + J_0^{kj} \hD{kl}{i} = 0.
\end{eqnarray}
We solve this relation for both degenerate and nondegenerate $J_0^{ij}$. 

When $J_0^{ij}$ is nondegenerate  we can immediately enforce $S^{ij}_{\phm i \phm j l}=0$ by setting 
\begin{equation} \label{eqn:inverse}
J_0^{kl} \omega_{lm} =   \de_{\phm k m}^k\,,
\end{equation}
and  $\hD{kl}{i}=0$.  In this case the quantity $\omega_{lm}$ denotes the components of the usual symplectic two-form dual to the cosymplectic form $J_0^{kl}$.

When $J_0^{ij}$ is degenerate, however, (\ref{eqn:inverse}) cannot be
solved, because $\de_{\phm k l}^{k}$ projects onto the null-space of
$J_0^{km}$.  Nevertheless,  $\omega_{ml}$ can be chosen as  a
generalized inverse \cite{RM71} of $J_0^{km}$, i.e.\  as solution of the
underdetermined system
\begin{equation} \label{eqn:geninv}
J_0^{kl} \omega_{lm} =   \de_{\phm k m}^k - \xif{\al}{k}\xiv{\al}{m} \,.
\end{equation}
Here, the covariant vectors
\begin{equation} \label{eqn:nullvect}
\xiv{\al}{l} := \dpar{C^{\al}(z_0)}{z_0^l},  
\end{equation}
span the null space of $J_0^{kl}$, while the contravariant vectors $\xif{\al}{k}$ are defined as  their duals, i.e.\   they satisfy the bi-orthogonality (pairing) relation
\begin{equation} \label{eqn:biorth}
\xif{\be}{l} \xiv{\al}{l} = \delta_\be^{\phm \be \al}.
\end{equation}
Observe, the right-hand-side of  (\ref{eqn:geninv}) no longer contains components in the null space of $J_0^{km}$ and so an inverse is possible. 

For $\hD{kl}{i}$, we choose
\begin{equation} \label{eqn:dhat}
\hD{kl}{i}=\frac{1}{3} \dpar{\xiv{\al}{l}}{z_0^k} \xif{\al}{i} + \A{\al}{k}{i} \xiv{\al}{l} + \A{\al}{l}{i} \xiv{\al}{k},
\end{equation} 
where the coefficients $\A{\al}{k}{i}$ will be determined later. Note that $\hD{kl}{i}$ is symmetric in $kl$ as required, because
\begin{equation} \label{eqn:updown}
\dpar{\xiv{\al}{l}}{z_0^k} = \dpar{^2 C^{\al}}{z_0^l \partial z_0^k}
             = \dpar{\xiv{\al}{k}}{z_0^l}.
\end{equation}
Introducing (\ref{eqn:geninv}) and (\ref{eqn:dhat}) in (\ref{eqn:compl}), and using (\ref{eqn:updown}),  
(\ref{eqn:nullvect})  and its consequence
\begin{equation} \label{eqn:dJdxi}
\dpar{J_0^{im}}{z_0^l} \xiv{\al}{m} = - J_0^{im} \dpar{\xiv{\al}{m}}{z_0^l}
\end{equation}
yields
\begin{equation} \label{eqn:toto}
\xiv{\al}{l} \left(
J_0^{ik} \A{\al}{k}{j} + J_0^{kj} \A{\al}{k}{i} -
\frac{1}{3} \dpar{J_0^{ij}}{z_0^k} \xif{\al}{k} \right) = 0.
\end{equation}
Using the same properties as previously, it can be checked that for $\A{\al}{k}{i}$ defined by
\begin{equation} \label{eqn:A}
\A{\al}{k}{i} = - \frac{1}{6} \dpar{J_0^{in}}{z_0^m} \omega_{nk} \xif{\al}{m} +
\frac{1}{6} \xif{\be}{i} \dpar{\xiv{\be}{k}}{z_0^m} \xif{\al}{m}
\end{equation}
the factor between parenthesis in (\ref{eqn:toto}) vanishes identically.

We have thus constructed a general solution of (\ref{eqn:condd}). Collecting
(\ref{eqn:decomp}), (\ref{eqn:dhat}), and (\ref{eqn:A}), we can write $\D{kl}{i}$ in the form:
\begin{equation} \label{eqn:D}
\D{kl}{i} = \frac{1}{3}\dpar{J_0^{in}}{z_0^m} \omega_{kn} \left(\de_l^m + \frac{1}{2} 
		\xif{\al}{m} \xiv{\al}{l} \right)
	+ \frac{1}{6} \xif{\be}{i} \dpar{\xiv{\be}{k}}{z_0^m} 
		\left(\de_l^m + \xif{\al}{m} \xiv{\al}{l} \right) 
	+ (k \leftrightarrow l),
\end{equation}
where $k \leftrightarrow l$ designates the symmetric term in $kl$.

So far, we have only considered the Poisson matrix, and not the
Hamiltonian. The change of variable (\ref{eqn:cv}) needs to be
inverted to compute the Hamiltonian $F_\eta$. To the required order
of accuracy, the inversion is immediate, and gives
\begin{equation} \label{eqn:cvi}
z^i = \eta^i - \frac{\ep}{2} \, \D{kl}{i} \, \eta^k \eta^l + O(\ep^2).
\end{equation}

\subsection{Lie-Poisson systems}
\label{ssec:lpsystems}

Typically dynamical systems that describe matter have Poisson matrices with  Lie-Poisson form, which in finite dimensions is given 
\[
J^{ij} = \stc{ij}{k} z^k,
\]
where the $\stc{ij}{k}$ are the structure constants of some Lie algebra.  In general,  such Lie-Poisson systems can be flattened by the procedure of Section \ref{ssec:flat}.  

In the case where the algebra with structure constants $\stc{ij}{k}$  is semisimple \cite{jac,gil},  the flattening transformation  can be found more directly.  For semisimple Lie algebras,   the   Cartan-Killing symmetric tensor, 
\[
g^{ij} := - \stc{ik}{l} \stc{lj}{k}
\]
has an ordinary  inverse $g_{ij}$, i.e. 
\[
g_{ik} \, g^{kj} = \de_i^{\phm i j}.
\]
The `metric' tensor $g^{ij}$ can then be used to put the structure constants
in a fully antisymmetric form $c^{ijk}$ that satisfies
\begin{equation} \label{eqn:lp1}
c^{ijk} = \stc{ij}{l}  g^{lk}.
\end{equation}
It can also be used to relate the covariant and contravariant vectors 
$\xiv{\al}{i}$ and $\xif{\al}{i}$ through an expression of the form
\begin{equation} \label{eqn:lp2}
\xif{\al}{i} = f_{\al \be} \, g^{ik} \xiv{\be}{k},
\end{equation}
for some symmetric tensor $f_{\al \be}$. Exploiting this expansion, one can find 
a direct solution of (\ref{eqn:toto}) given by
\begin{equation} \label{eqn:lp3}
\A{\al}{k}{i}=- \frac{1}{6} f_{\al \be}\,  g^{im} \dpar{\xiv{\be}{k}}{z_0^m}.
\end{equation}
Indeed, using (\ref{eqn:lp1})--(\ref{eqn:lp3}), (\ref{eqn:dJdxi}) and 
\[
\dpar{J_0^{im}}{z_0^l} = \stc{im}{l},
\]
one can find
\begin{eqnarray*}
J_0^{ik} \A{\al}{k}{j} + J_0^{kj} \A{\al}{k}{i} & = & 
    \frac{1}{6} \left( \stc{ik}{m} g^{jm} 
                     + \stc{kj}{m} g^{im} \right) 
                              f_{\al \be}\,  \xiv{\be}{k} \\
&=& - \frac{1}{3} c^{ijk} f_{\al \be} \, \xiv{\be}{k} 
=  - \frac{1}{3} c^{ijk} g_{lk} \xif{\al}{l}  \\
&=& - \frac{1}{3} \stc{ij}{l}  \xif{\al}{l} 
= - \frac{1}{3} \dpar{J_0^{ij}}{z_0^l} \xif{\al}{l},
\end{eqnarray*}
so that (\ref{eqn:toto}) is satisfied. Therefore, the change of variable
(\ref{eqn:cv}) is defined by
\begin{equation} \label{eqn:Dlp}
\D{kl}{i} = \frac13 \stc{im}{l} \omega_{km} + \frac{1}{6} f_{\al \be} \,
g^{im} \left( \xiv{\be}{m} \dpar{\xiv{\al}{k}}{z_0^l} 
             -\xiv{\al}{l} \dpar{\xiv{\be}{k}}{z_0^m} \right) \, +
(k \leftrightarrow l).
\end{equation}


Semisimple Lie-Poisson brackets possess a quadratic Casimir invariant given by
\begin{equation} \label{eqn:qCas}
C= g_{ij} z^i z^j\,,
\end{equation}
which we record here for later use.

\section{Applications} \label{sec:app}

\subsection{Three-dimensional semisimple Lie-Poisson brackets}

It is well-known that there are nine real Lie algebras of dimension three \cite{jac}.  In this subsection we consider the Lie-Poisson brackets associated with two of these, $\mathfrak{so}(3)$ and $\mathfrak{so}(2,1)$, both of which are semisimple.   The bracket determined by $\mathfrak{so}(3)$  describes, e.g.,  spin systems and  Euler's equations for the free rigid body \cite{SM}, while  $\mathfrak{so}(2,1)$ emerges naturally from quadratic moment projections of Euler's fluid equations  and describes, e.g.,    Kida vortex dynamics \cite{NMM96,MMF97} .

The rotation algebra $\mathfrak{so}(3)$ has structure constants $c^{ij}_k=\eps_{ijk}$, with the   $\eps_{ijk}$ being the purely antisymmetric Levi-Civita tensor, while the structure constants of  $\mathfrak{so}(2,1)$ are the same except for a sign flip.  Both  Poisson matrices are expressed by the  following:
\begin{equation} \label{eqn:j0euler}
J_0^{\pm}= \left(
\begin{array}{ccc}
   0    & z_0^3 &  - z_0^2 \\
- z_0^3 &    0  & \pm z_0^1 \\
   z_0^2 & \mp z_0^1 & 0
\end{array}
\right)\,,
\end{equation}
where the upper sign corresponds to $\mathfrak{so}(3)$  and the lower to $\mathfrak{so}(2,1)$. 
 
For these semisimple algebras we can identify the algebras with there duals by their Cartan-Killing forms, which we scale as $g^{ij} = \mp 2 \delta^{ij}_\pm$ to give the following metrics:
\begin{equation}
\label{eqn:metrics}
\delta^{\pm}=  \left(
\begin{array}{ccc}
 \pm 1   & 0 & 0  \\
0 &    1 & 0 \\
0 & 0 & 1
\end{array}
\right)\,,
\end{equation}
where again  the upper sign corresponds to $\mathfrak{so}(3)$  and the lower to $\mathfrak{so}(2,1)$.

 In terms of $\de^{\pm}$ the structure constants for both cases can be represented as 
\[
c^{ij}_k=\eps^{ijs}\de^{\pm}_{sk}\,.
\]
Using (\ref{eqn:metrics}) the quadratic Casimirs take the form 
\[
|z_0|_{\pm}^2:=  \de^{\pm}_{ij} z_0^iz_0^j =:\langle z_0,z_0\rangle_{\pm}
\]
with the corresponding null eigenvectors and their duals  given by
\begin{equation}
\label{eqn:nulls}
\chi_{i} = \frac{\delta^{\pm}_{ij} z_0^j}{|z_0|_{\pm}} 
\qquad\mathrm{and}\qquad
\chi^{i} = \frac{z_0^i}{|z_0|_{\pm}}\,.
\end{equation}

Solving (\ref{eqn:geninv}), one finds that the generalized inverses of the two matrices are given by 
\begin{equation}
\label{eqn:omepm}
\omega^{\pm} = \frac{1}{|z_0|^2_{\pm}} \left(
\begin{array}{ccc}
   0    &  - z_0^3 &  z_0^2 \\
   z_0^3 &    0  &  - z_0^1 \\
   - z_0^2 &   z_0^1 & 0
\end{array}
\right)\,,
\end{equation}
which can easily be checked by substituting (\ref{eqn:omepm})  into (\ref{eqn:geninv}) and making use of (\ref{eqn:nulls}). 
Observe $\omega_{ij}^{\pm}= -\epsilon_{ijk}z^k/|z_0|^2_{\pm}$.

The different terms of (\ref{eqn:Dlp}) can now be evaluated  for  $\mathfrak{so}(3)$ and $\mathfrak{so}(2,1)$ together,  
\begin{eqnarray*}
\frac13 \stc{im}{l} \omega_{km} &=& \frac{z_0^r }{3 |z_0|_{\pm}^2} \eps^{ims} \delta^{\pm}_{s l}\,  \eps_{mkr} 
  =  \frac{1}{3 |z_0|_{\pm}^2} \big(z_0^i \de^{\pm}_{kl} - z_0^r \de_{k}^i \delta^{\pm}_{rl}\big), \\
\frac{1}{6} f_{\al \be} \,
g^{im} \left( \xiv{\al}{m} \dpar{\xiv{\be}{k}}{z_0^l} 
             -\xiv{\al}{l} \dpar{\xiv{\be}{k}}{z_0^m} \right) 
& = & \frac{z_0^r}{6 |z_0|_{\pm}^2} \de_{\pm}^{im} \big(\delta^{\pm}_{mr} \de^{\pm}_{kl} - \delta^{\pm}_{lr} \de^{\pm}_{km}\big) 
= \frac{1}{6 |z_0|_{\pm}^2} \big(z_0^i \de^{\pm}_{kl} - z_0^r \de_{ik}  \de^{\pm}_{lr} \big)\,,
\end{eqnarray*}
where  $f_{\pm} =\mp 1/2$ follows from $\chi^i= f_{\pm} g_{\pm}^{ik} \chi_k$.  Combining these results leads to
\[
\D{kl}{i} = \frac{1}{|z_0|_{\pm}^2}\left[ 
z_0^i \de^{\pm}_{kl} - \frac{z_0^r}{2} \big( \de^i_{k} \de^{\pm}_{lr} +  \de^i_{l} \de^{\pm}_{kr}  \big)
\right]\,,
\]
and thus to the new variable
\begin{equation} \label{eqn:etarb}
\eta^i = z^i + \frac{\ep}{2 |z_0|_{\pm}^2}  \big(
z_0^i \, |z|_{\pm}^2 - z^i \langle z,  z_0\rangle_{\pm}
\big)\,.
\end{equation}
Note that the change of variable is such that
\begin{equation} \label{eqn:norm}
|\eta|_{\pm}^2 = |z|_{\pm}^2 + O(\ep^2).
\end{equation}
With the variable $\eta$, the weakly nonlinear dynamics of the
disturbance is generated by the Poisson matrix  (\ref{eqn:j0euler}). The
corresponding Casimir function is
\begin{equation} \label{eqn:aaa}
\langle z_0, \eta \rangle_{\pm} = \langle z_0, z\rangle_{\pm} + \frac{\ep}{2} |z|_{\pm}^2 - \frac{\ep}{2
|z_0|_{\pm}^2} \langle z_0,  z\rangle_{\pm}^2.
\end{equation}
The first two terms can be recognized as the exact
Casimir
$\ep^{-1} (|z_0 + \ep z|_{\pm}^2 - |z_0|_{\pm}^2)/2$. The time derivative of the
third term is $O(\ep^2)$, so that it appears constant to the order of
accuracy considered here.

\subsection{Heavy-top bracket}

Another simple bracket is that of the heavy top for which the
dynamical variables consist of the angular momentum $\mu$ and the
direction of the gravity field $\ga$: $z=(\mu,\ga)\tr$. The
Poisson matrix defining this bracket can be written
\[
J^{ij} = \left(
	\begin{array}{cc} 
-\eps_{ijk} \mu^k & -\eps_{ijk} \ga^k \\
-\eps_{ijk} \ga^k & 0 
	\end{array}
	 \right).
\]
The heavy top bracket \cite{SM} has a Lie-Poisson
structure based on a Lie algebra extension of $\mathfrak{so}(3)$, but the corresponding Lie algebra is not semisimple \cite{TM00}; we
will thus use (\ref{eqn:D}) to find the variable transformation
(\ref{eqn:cv}). For simplicity, we restrict our attention to
equilibrium solutions 
with $\mu=(0,0,M_3)$ and $\ga=(0,0,1)$ which arise for the Lagrange
top (top with a symmetry). The following tensors are required for
(\ref{eqn:D}) and
can be easily obtained:
\[
J_0^{ij} = \left( \begin{array}{cc}
	- M_3 \eps_{ij3} & - \eps_{ij3} \\
	- \eps_{ij3}       &     0
	          \end{array} \right),
\
\xiv{1}{i} = \left( \begin{array}{c} 
	  \de_{i3} \\
	     0
	             \end{array} \right),
\
\xiv{2}{i} = \left( \begin{array}{c} 
	  	0  \\
	     \de_{i3}
	             \end{array} \right),
\
\omega_{nk} =-\frac{1}{3} \left( \begin{array}{cc}
	          0        &  \eps_{nk3} \\
	 \eps_{nk3}        & -M_3 \eps_{nk3}
	          \end{array} \right).
\]
Because the null eigenvectors are orthonormal, we have taken
$\xif{\al}{i}=\xiv{\al}{i}$. A convenient way to denote tensors such
as $\D{kl}{i}$ is to make the separation $(\D{kl}{\bi},\D{kl}{\ti})$
between the components $i$ corresponding to $\mu$ (denoted by $\bi$)
and those corresponding 
to $\ga$ (denoted by $\ti$). With this convention, we can write
$\partial J_0^{in} / \partial z_0^m$ as the two  triplets of $6 \times
6$ matrices 
\[
\dpar{J_0^{\bi n}}{z_0^m} = \left( \begin{array}{cc}
		- \eps_{\bi nm} & 0 \\
	        0 & - \eps_{\bi nm}  
			           \end{array} \right),
\quad 
 \dpar{J_0^{\ti n}}{z_0^m} = \left( \begin{array}{cc}
		0 & -\eps_{\ti n m} \\
	        0 & 0  
			           \end{array} \right), \quad \bi,\ti=1,2,3, 
\]
where $n$ ($m$) is the row (column) index. Noting that the general
expression for two orthonormal null eigenvectors of $J^{ij}$ is
\[
\xiv{1}{k}=\frac{1}{N} (\ga^k,\mu^k - (\mu \cdot \ga)/|\ga|^2
\ga^k)\tr, \quad \xiv{2}{k}=\frac{1}{|\ga|}(0,\ga^k)\tr,
\]
where $N^2:= |\mu|^2 + |\ga|^2 - (\mu \cdot \ga)^2/|\ga|^2$, one can
calculate the two matrices
\begin{eqnarray*}
\dpar{\xiv{1}{k}}{z_0^m} &=& \left( \begin{array}{cc}
		0       &  \de_{km} - \de_{k3} \de_{m3} \\
 \de_{km} - \de_{k3} \de_{m3}   &  - M_3 ( \de_{km} - \de_{k3}
		\de_{m3})
				\end{array}
	\right) \\
\dpar{\xiv{2}{k}}{z_0^m} &=& \left( \begin{array}{cc}
		0       &  0 \\
                0   &   \de_{km} - \de_{k3} \de_{m3})
				\end{array}
	\right)
\end{eqnarray*}
With these results, and using the skew-symmetry of $\omega_{nk}$, one can
compute (\ref{eqn:D}) through simple matrix multiplications and
additions. After simplifications, one finds
\begin{eqnarray*}
\D{kl}{\bi} &=& \left( \begin{array}{cc}
		0       &  \de_{kl} \de_{\bi 3} - {\scriptstyle
		\frac{1}{2}} (\de_{\bi k} \de_{l3} + \de_{\bi l} \de_{k3}) \\
 \de_{kl} \de_{\bi 3} - {\scriptstyle \frac{1}{2}} 
(\de_{\bi k} \de_{l3} + \de_{\bi l} \de_{k3})  &
- M_3 [\de_{kl} \de_{\bi 3} - {\scriptstyle
		\frac{1}{2}} (\de_{\bi k} \de_{l3} + \de_{\bi l} \de_{k3})]
				\end{array}
	\right) \\
\D{kl}{\ti} &=& \left( \begin{array}{cc}
		0       &  0 \\
                0       &
\de_{kl} \de_{\ti 3} - {\scriptstyle
		\frac{1}{2}} (\de_{\ti k} \de_{l3} + \de_{\ti l} \de_{k3})
				\end{array} \right).
\end{eqnarray*}
The new variables which can be used to derive the Hamiltonian
equations for the weakly  nonlinear dynamics of the heavy top take
thus the form
\[
\left\{
\begin{array}{l}
\eta^\al = \mu^1 - \displaystyle{\frac{\ep}{2}} \left(\mu^\al \ga^3 +
\mu^3 \ga^\al - M_3 \ga^\al 
\ga^3\right) \smallskip \\
\eta^3 = \mu^3 + \displaystyle{\frac{\ep}{2}} \left\{\mu^1 \ga^1 +
\mu^2 \ga^2 - M_3 
\left[(\ga^1)^2+(\ga^2)^2\right]\right\} \smallskip \\
\si^\al = \ga^1 - \displaystyle{\frac{\ep}{2}} \ga^\al \ga^3  \smallskip\\
\si^3 = \si^3 + \displaystyle{\frac{\ep}{2}} \left[(\ga^1)^2 + (\ga^2)^2\right]
\end{array}
\right. \quad \al=1,2
\]

\subsection{Two-spin system} \label{sec:twospin}

To illustrate the interest of the approach, we now consider the specific example of a system constituted of two coupled spins in a magnetic field.
This system is defined by the Hamiltonian
\[
H = B \cdot \om + B \cdot \Om + a \, \om \cdot \Om,
\]
where $B$ is the external magnetic field, $a$ a coupling coefficient,
and $\om$ and $\Om$ the angular momentum vectors of the two spins, and
by the bracket
\begin{equation} \label{eqn:pbnlin}
\{f,g\} = \sum_{\al=1}^{2} \eps_{ijk} \, \oma^k \dpar{f}{\oma^i} \dpar{g}{\oma^j}
.
\end{equation}
i.e.\  the sum of two rigid body brackets corresponding to each spin. 
The dynamical equations are
\[
\dot{\omega}_{(\al)} = B \times \oma + a \, \omega_{(\al+1)} \times \oma,
\qquad \al=1,2. 
\]
(The system is integrable, because $\om \cdot \Om$ is conserved in
addition to the energy and two Casimirs.)
An obvious equilibrium solution of these equations is given by 
\begin{equation} \label{eqn:2spineq}
{\om}_0 = {\Om}_0 =
m B. 
\end{equation}
Noting that the bracket possesses two Casimir functions
\[
C^{\al} = |\oma|^2, \quad \al = 1,2,
\]
it is easily seen that this equilibrium is a critical point of
\[
F = H + \lambda_{1} C^{1} + \lambda_{2} C^{2}
\]
with
\[
 \lambda_{1} = \lambda_{2} = \la := -\frac{1 + a m}{2m}.
\]
We can choose one coordinate axis aligned with the external magnetic field, so that $B=(B_1,0,0)$ and ${\om}_0 = {\Om}_0 = (mB_1,0,0)$. 
We now study the weakly nonlinear evolution of a small disturbance 
to this equilibrium. We therefore introduce the decomposition
\[
{\oma} = {\oma}_0 + \ep \za, \quad \al=1,2, \qquad \mbox{with } \ep \ll 1.
\]
In terms of the perturbation $\za$, the free energy (\ref{eqn:freeen})
is given by 
\begin{equation} \label{eqn:hz}
F_z := \ep^{-2} \left[ F({\oma}_0 + \ep \za) - F({\oma}_0) \right]
  = a \z \cdot \Z + \la \left( |\z|^2 + |\Z|^2 \right),
\end{equation}
which is exactly quadratic. The nonlinearity in the equation governing
the evolution of $\za$ comes thus entirely from the bracket. Note that the stability of the equilibrium can be tested using
Dirichlet's
criterion.  The matrix
\[
\frac{1}{2} \dpar{^2 F_z}{\z \partial \Z} = \left(
\begin{array}{c c}
\la & a/2 \\
a/2 & \la
\end{array}
\right)
\]
has the eigenvalues $-(1 + 2 am)/2m$ and $-1/2m$ and is thus sign-definite if $1 + 2 am
> 0$. This condition ensures the nonlinear stability of the equilibrium.

Since the nonlinear bracket is the combination of two rigid body
brackets, we can apply the change of variable (\ref{eqn:etarb}) with $z_0 =
(mB_1,0,0)^\mathrm{T}$ and
define
\begin{equation} \label{eqn:etaz}
\left\{
	\begin{array}{l}
		\etaa^1 = \za^1 + \displaystyle{\frac{\ep}{2 mB_1}}
			\left[(\za^2)^2  +
			(\za^3)^2\right] \smallskip \\
		\etaa^2 = \za^2 - \displaystyle{\frac{\ep}{2 mB_1}}
			\za^1 \za^2 
			 \smallskip \\
		\etaa^3 = \za^3 - \displaystyle{\frac{\ep}{2 mB_1}}
			 \za^1 \za^3
	\end{array}
\right. \al=1,2.
\end{equation}
In accordance with (\ref{eqn:cond}), the dynamics of $\etaa$ including
the $O(\ep)$ nonlinear terms is determined by the constant bracket
\begin{equation} \label{eqn:pblin0}
\{f,g\}_0 = m B_1 \sum_{\al=1}^{2} \eps_{ij1}  \dpar{f}{\etaa^i} \dpar{g}{\etaa^j}.
\end{equation}
It is therefore immediately clear that 
\begin{equation} \label{eqn:c1c2}
\dot{\eta}_{(1)}^1=\dot{\eta}_{(2)}^1=0, \quad \mbox{hence} \quad \eta^1_{(1)}
= c_1, \: \eta^1_{(2)} = c_2,
\end{equation}
where $c_1, c_2$ are fixed by the initial conditions. In fact,
$\eta^1_{(1)}$ and $\eta^1_{(2)}$
are the Casimir functions of the system.

The nonlinearity in the evolution equations for $\etaa$ appears as cubic terms in the
Hamiltonian.
To compute those terms, we need to invert the transformation
(\ref{eqn:etaz}) according to
\begin{equation} \label{eqn:zeta}
\left\{
	\begin{array}{l}
		\za^1 = \etaa^1 - \displaystyle{\frac{\ep}{2 mB_1}}
			\left[(\etaa^2)^2  +
			(\etaa^3)^2\right] + O(\ep^2) \smallskip \\
		\za^2 = \etaa^2 + \displaystyle{\frac{\ep}{2 mB_1}}
			\etaa^1 \etaa^2 + O(\ep^2)
			 \smallskip \\
		\za^3 = \etaa^3 + \displaystyle{\frac{\ep}{2 mB_1}}
			 \etaa^1 \etaa^3 + O(\ep^2)
	\end{array}
\right. \al=1,2.
\end{equation}
Introducing this into the Hamiltonian (\ref{eqn:hz}) and using (\ref{eqn:norm}) yields
\begin{equation} \label{eqn:titi}
F_\eta = a \eta_{(1)} \cdot \eta_{(2)}
 + \la \left( |\eta_{(1)}|^2 + |\eta_{(2)}|^2 \right) + \frac{\ep
 a}{2mB_1} \sum_{i=2}^3 \left( \eta_{(1)}^i - \eta_{(2)}^i \right)
 \left(\eta_{(1)}^1 
 \eta_{(2)}^i - \eta_{(2)}^1 \eta_{(1)}^i \right) + O(\ep^2).
\end{equation}

The weakly
nonlinear equations for $\eta := (\eta_{(1)}^2,\eta_{(1)}^3,\eta_{(2)}^2,\eta_{(2)}^3)^{\text{T}}$ can then be rendered canonical. By construction, the necessary variable transformation uses the action--angle coordinates  $(J_{(\alpha)},\theta_{(\alpha)})$ which make the linearized equations for $z$ canonical; these coordinates are derived in Appendix \ref{app:twospin}. Applied to $\eta_{\alpha}$, they are defined by
\begin{equation} \label{eqn:ejt}
\left\{
	\begin{array}{l}
\eta_{(1)}^2 = \sqrt{|mB_1|} \left[ - \sqrt{\smash[b]{\j}} \sin(s \th) + \sqrt{\smash[b]{\jj}}
\sin(s \Th) \right] \nonumber \\
\eta_{(1)}^3 = \sqrt{|mB_1|} \left[ - \sqrt{\smash[b]{\j}} \cos(s \th) + \sqrt{\smash[b]{\jj}}
\cos(s \Th) \right] \nonumber \\
\eta_{(2)}^2 = \sqrt{|mB_1|} \left[  \sqrt{\smash[b]{\j}} \sin(s \th) + \sqrt{\smash[b]{\jj}}
\sin(s \Th) \right] \nonumber \\
\eta_{(2)}^3 = \sqrt{|mB_1|} \left[  \sqrt{\smash[b]{\j}} \cos(s \th) + \sqrt{\smash[b]{\jj}}
\cos(s \Th) \right]
	\end{array}
\right. ,
\end{equation}
where $s:=\textrm{sign}(mB_{1})$.
In terms of $(J_{(\alpha)},\theta_{(\alpha)})$, the Hamiltonian (\ref{eqn:titi})  becomes 
\begin{eqnarray*}
F_{(\Ja,\tha)} &=& - s  \sum_{\al=1}^2 \si_{(\al)} \Ja + a c_1 c_2 \\
&-& \ep s a
\sqrt{\smash[b]{\j}} \left[ (c_1+c_2) \sqrt{\smash[b]{\j}}  + (c_1 - c_2)
\sqrt{\smash[b]{\jj}} 
\cos(\th-\Th) \right] + O(\ep^2),
\end{eqnarray*}
where
\begin{equation} \label{eqn:unpfre}
\si_{(1)} :=  B_1 (1+2 am) \quad \textrm{and} \quad
\si_{(2)} := B_1.
\end{equation}
This completes the reformulation of the weakly nonlinear dynamics of the two-spin system as a canonical system to which standard pertubation methods can be applied. As an immediate illustration, we can read off from (\ref{eqn:ejt}) the frequencies  $\si_{(1)} + \ep a (c_1+c_2)$ and $\si_{(2)}$ and interpret the correction $\ep a (c_1+c_2)$ to $\si_{(1)}$ as a nonlinear frequency shift. Further details on the perturbation theory are given in Appendix \ref{app:twospin}.

\section{Infinite-dimensional systems} \label{sec:infinite}

The extension of (\ref{eqn:hamilton}) to partial differential equations is
\begin{equation} \label{eqn:hamilton2}
\dpar{\zeta}{t} = \J \dfun{\H}{\zeta},
\end{equation}
where $\zeta$ is the vector of the dynamical variables, $\J$ is a
skew-adjoint operator that we call the Poisson operator,  and the Hamiltonian $\H$ is a functional
\[
\H = \int H(x,t,\zeta, \nabla \zeta, \dots) \dif \x\,.   
\]
The Jacobi identity satisfied by $\J$ is best written in terms of the
Poisson bracket 
\[
\{\F,\G\} := \int \dfun{\F}{\zeta} \J(\zeta) \dfun{\G}{\zeta} \dif \x
\]
as
\[
\{\F,\{\G,\H\}\} + \{\G,\{\H,\F\}\} + \{\H,\{\F,\G\}\} = 0.
\]
As in the finite-dimensional case, the operator $\J$ can be degenerate
and there exists Casimir functionals $\C$ such that
\[
\{\F,\C\}=0, \qquad \forall \F.
\]

Consider an equilibrium $\zeta_0$ of (\ref{eqn:hamilton2}).  Usually only  trivial equilibria are 
 critical points of $\H$, while equilibria of interest are  critical points of the combination 
\[
\F[\zeta] := \H[\zeta] + \C[\zeta]
\]
for a well-chosen Casimir functional, i.e.
\begin{equation} \label{eqn:critic}
\J_0 \dfun{\F_0}{\zeta_0}:=\J(\zeta_0) \dfun{\F[\zeta_0]}{\zeta_0} = 0.
\end{equation}
To study the evolution of a small disturbance to $\zeta_0$, we introduce
the decomposition
\begin{equation} \label{eqn:decomp2}
\zeta = \zeta_0 + \ep \zeta^\prime
\end{equation}
and omit the prime. The disturbance obeys the equation
\[
\dpar{\zeta}{t} = \J_\zeta \dfun{\F_\zeta}{\zeta},
\]
where 
\begin{equation} \label{eqn:dist2}
\J_\zeta := \J(\zeta_0 + \ep \zeta) \quad \text{ and } \quad \F_\zeta := \ep^{-2}
[F(\zeta_0 + \ep \zeta)-F(\zeta_0)].
\end{equation}
Because of (\ref{eqn:critic}), the linearized equation given by
\[
\dpar{\zeta}{t} = \J_0 \left. \dfun{^2 \F_\zeta}{\zeta^2} \right|_{\zeta=0} \zeta
\]
is Hamiltonian. An equation describing the weakly nonlinear evolution
of the disturbance can be obtained by truncating (\ref{eqn:dist2}) to
$O(\ep)$. Generally, it will not be Hamiltonian, because of the
truncation of $\J_\zeta$. To find a truncation which is Hamiltonian, we
follow the same idea as for finite-dimensional systems: we seek a near-identity
transformation from $\zeta$ to $\eta$ such that
\begin{equation} \label{eqn:cond2}
\J_\eta = \J_0 + O(\ep^2),
\end{equation}
where $\J_\eta$ is the Poisson operator  for $\eta$. The weakly
nonlinear evolution can therefore be determined from the Hamiltonian
equation
\[
\dpar{\eta}{t} = \J_0 \dfun{\F_\eta}{\eta} + O(\ep^2),
\]
where $\F_\eta := \F_\zeta[\zeta(\eta)]$ can be truncated to
$O(\ep)$. We emphasize that this procedure does not modify the
independent variables used; this is important for practical use of
the weakly nonlinear equations (cf. Lagrangian vs. Eulerian description). 

The required change of variable has the general form
\begin{equation} \label{eqn:CV2}
\eta = \zeta + \ep \O(\zeta;\zeta),
\end{equation}
where $\O$ is a symmetric bilinear operator to be determined. From this, the operator
$\J_\eta$, defined by
\begin{equation} \label{eqn:tata}
\{\F,\G\}=\int \dfun{\F}{\eta} \J_\eta \dfun{\G}{\eta} \dif \x 
= \int \dfun{\F}{\zeta} \J_\zeta \dfun{\G}{\zeta} \dif \x,
\end{equation}
can be obtained, and an equation for $\O$ can be found from condition
(\ref{eqn:cond2}). Although this procedure is general, it is clear in
 light of the finite-dimensional case that the algebra involved can be
quite complex. Therefore, we shall consider particular examples
to show how the change of variable can be found and exploited. Our
examples concern two different brackets and (at least)
four different physical systems: (a) Euler bracket \cite{mor82,olv82,MW83}, which appears for
the two-dimensional scalar vortex dynamics (including quasi-geostrophy) and for
the one-dimensional Vlasov--Poisson equation \cite{mor80}, (b) the bracket which appears
for two-dimensional stratified fluid and  for two-dimensional MHD \cite{MH84,MM84}. In
all cases, we will consider simple equlibria,
depending on one coordinate only.

\section{Euler bracket, vortex dynamics and Vlasov--Poisson equations}
 \label{sec:eul}

The Euler bracket in two dimensions $(x,y)$ can be written
\begin{equation}  \label{eqn:breuler}
\{\F,\G\} = \int \!\! \int \zeta
\left[\dfun{\F}{\zeta},\dfun{\G}{\zeta}\right] \dif x \text{d}y, 
\end{equation}
where
\[
[f,g] := \dpar{f}{x} \dpar{g}{y} - \dpar{f}{y} \dpar{g}{x},
\]
and corresponds to the Poisson operator
\[
\J(\zeta)=-[\zeta,\cdot].
\]
The equilibrium solution is assumed to be a function of $y$ only,
$\zeta_0=\zeta_0(y)$, and therefore,
\begin{equation} \label{eqn:gardner}
\J_0 = \zetaop \dpar{}{x},
\end{equation}
with $\zetaop := \mathrm{d} \zeta_0/\mathrm{d} y$.
We consider a simple form of the change of variable (\ref{eqn:CV2})
given by 
\begin{equation} \label{eqn:CV3}
\eta = \zeta + \frac{\ep}{2} \DD \zeta^2,
\end{equation}
where the linear operator $\DD=\DD(\zeta_0)$ can be expected to involve
$\partial_y$ and not $\partial_x$. The functional derivatives with
respect to $\zeta$ and $\eta$ are related by
\[
\dfun{F}{\zeta} = \left( 1 + \ep \zeta \DD^\dagger \right)
\dfun{F}{\eta},
\]
where $^\dagger$ denotes the adjoint. 
From (\ref{eqn:tata}) it is then found that
\begin{eqnarray*}
\int \!\! \int \dfun{\F}{\zeta} \J_\eta \dfun{\G}{\zeta} \dif x \text{d}y &=&
 \int \!\! \int \left\{ \dfun{\F}{\zeta} \J_0 \dfun{\G}{\zeta} 
	+ \ep \zeta \left(\zetaop \DD^\dagger \left(
\dfun{\F}{\zeta} \right) \partial_x
\left(\dfun{\G}{\zeta}\right) \right. \right. \\ 
&-& \left. \left. \zetaop \partial_x
\left(\dfun{\F}{\zeta}\right) 
\DD^\dagger 
\left(\dfun{\G}{\zeta} \right)+ \left[\dfun{\F}{\zeta},\dfun{\G}{\zeta}\right] \right) \right\}
\dif x \text{d}y + O(\ep^2),
\end{eqnarray*}
Clearly, condition (\ref{eqn:cond2}) is satisfied if
\[
\DD^\dagger = \frac{1}{\zetaop} \dpar{}{y} \quad \textrm{i.e.} \quad \DD =
- \dpar{}{y} \frac{1}{\zetaop}.
\]
Hence, the change of variable sought is
\begin{equation} \label{eqn:CV4}
\eta = \zeta - \ep \dpar{}{y} \left( \frac{\zeta^2}{2 \zetaop} \right).
\end{equation}
For now, let us assume that $\zetaop \not= 0$. 
The inverse transformation is  
given by
\begin{equation} \label{eqn:CV4inv}
\zeta = \eta + \ep \dpar{}{y} \left( \frac{\eta^2}{2 \zetaop} \right) +
O(\ep^2).
\end{equation}
Note that the use of $\eta$ requires not only $\ep \ll 1$, but also
\begin{equation} \label{eqn:wnl}
\ep \dpar{\zeta}{y} \ll \zetaop
\end{equation}
(and similar conditions for higher-order derivatives).

We make a few remarks to relate the transformation
(\ref{eqn:CV4})--(\ref{eqn:CV4inv}) to previous
results. 
\begin{enumerate}
\item It is clear from the bracket (\ref{eqn:gardner}) for
$\eta$ that
\begin{equation} \label{eqn:nomean}
\dpar{\overline{\eta}}{t}=O(\ep^2),
\end{equation}
where $^{\displaystyle{-}}$ denotes the average in the $x$ direction;
therefore, 
\begin{equation} \label{eqn:mean}
\overline{\zeta} = \ep \dpar{}{y} \left( \frac{\overline{\zeta^2}}{2
\zetaop} \right) + C(y) + O(\ep^2) 
                 = \ep \dpar{}{y} \left( \frac{\overline{\hat{\zeta}^2}}{2
\zetaop} \right) + C(y) + O(\ep^2), 
\end{equation}
where $\hat{\zeta}:= \zeta - \overline{\zeta}$ is the wavy part of
the disturbance, and $C(y)$ is fixed by the initial conditions.
That the mean value of $\zeta$ can be evaluated to leading order
knowing the wavy part of the disturbance only is a standard result of
wave--mean flow interaction theory for Euler or quasi-geostrophic
equations  (e.g.\ \cite{buhl09}).
It is often formulated using the velocity in the $x$ direction defined
by
\[
u := \int \zeta \dif y
\]
as
\[
\dpar{\overline{u}}{t}=\ep  \frac{\overline{\hat{\zeta}^2}}{2
\zetaop} + O(\ep^2),
\]
and the quantity $\zeta^2/2 \zeta^\prime_0$ is often refered to as the pseudomomentum (up to sign) \cite{shep90}, but can be traced back to the physical monentum of the fluid \cite{BM02}.  Here, we relate this result to the 
Poisson bracket, independently of the Hamiltonian. 

\item The approximate constancy of $\overline{\eta}$, which results from the
imposed Poisson bracket, suggests that $\eta$ must be given by the old
variable minus their $x$-average (compare (\ref{eqn:CV4}) and
(\ref{eqn:mean})). This yields a procedure for
constructing the change of variable if we can compute a priori this
average, i.e.\ the right-hand side of
(\ref{eqn:mean}). This is possible using the concept of dynamical
accessibility \cite{MP89}. This expresses the fact the perturbed vorticity $\zeta_{0} + \ep \zeta$ lies on the same symplectic leaf as the equilibrium vorticity $\zeta_{0}$, that is, the values of the Casimir functionals for the perturbed flows are the same as the equilibrium values. This can generally be arranged by an appropriate definition of the equilibrium flow. 
The dynamical accessibility of $\zeta_{0} + \eps \zeta$ can be written explicitly in the form 
\begin{equation} \label{eqn:dynacc}
\ep \zeta = \left(\exp\{-\G,\cdot\} - 1 \right) \zeta_0 
	  = \left(\exp[\ep g,\cdot] -1 \right) \zeta_0 = \ep g_x
	  \zetaop + \frac{1}{2} \ep^2 [g, g_x \zetaop] +
	  \ldots,
\end{equation}
where $g$ is an arbitrary function and
\[
\G := \ep \iint  g \zeta \dif x \text{d}y.
\]
Applying $^{\displaystyle{-}}$ and noting that $g_x = \zeta / \zetaop
+ O(\ep)$, one finds (\ref{eqn:mean}) and hence an indication of the
change of variable (\ref{eqn:CV4}). Note that the presence/absence of
$^{\displaystyle{-}}$ is not really relevant, as one can add a term in
$\partial_x$ to (\ref{eqn:CV4}) without changing the bracket to the
order of accuracy considered.

\item The dynamical accessibility condition (\ref{eqn:dynacc}) also make clear that the variable transformation (\ref{eqn:CV4}) is in fact well defined even when $\zetaop = 0$ at some $y$.

\item The transformation (\ref{eqn:CV4})  is the related Zakharov--Piterbarg variable transformation \cite{ZP88} for the
quasi-geostrophic system. Considering the particular case $\zeta_0 = \beta
y$, they found that the new variable implicitly defined by
\begin{equation} \label{eqn:zp}
\etazp(x,y) = \zeta \left[ x,y-\ep \beta^{-1} \etazp(x,y) \right]
\end{equation}
is such that the Poisson operator  is given by
\[
\J_{\etazp} = \zetaop \dpar{}{x} = \beta \dpar{}{x}
\]
to all orders in $\ep$. Expanding (\ref{eqn:zp}) in $\ep$ and comparing with
(\ref{eqn:CV4}) it is clear that our transformation is the  first-order
approximation to that of Zakharov and Piterbarg. In fact, their
transformation can be extended to more general forms of $\zeta_0$.
Suppose that $\zeta_0(y)$ is monotonic and admits the inverse
$y_0(\zeta)$, and consider the transformation of both dependent and
independent variables
$(x,y,\zeta) \rightarrow (x,z,\etac)$ defined by
\begin{equation} \label{eqn:etac}
z = y_0\left[\zeta_0(y) + \ep \zeta(x,y)\right] \quad \mbox{ and }
\quad \etac = \etac[x,y,\zeta] =\ep^{-1} 
\left.\dt{\zeta_0}{y_0}\right|_z (z - y).
\end{equation}
Assuming that this transformation is one to one (level sets of $\zeta_0 + \ep \zeta$ remain graphs over $x$), it can be shown 
that the Poisson operator for $\etac$ is exactly given
by (\ref{eqn:gardner}). This is detailed in Appendix \ref{app:zp}, where we also
show that $\etac$ reduce to $\etazp$ for $\zeta_0=\beta y$ and that
$\eta$ defined by (\ref{eqn:CV4}) is simply the $O(\ep)$ 
approximation to $\etac$.
It is interesting to note the ``Lagrangian character'' of the
transformed variables: $z$ is in fact a vorticity (transformed in a
space coordinate through the function $y_0$), hence a label, and
$\etac$ is related to a displacement.

\end{enumerate}

Below, we consider the vortex dynamics equation and the one-dimensional
Vlasov--Poisson equation and use the inverse transformation
(\ref{eqn:CV4inv}) to
compute the Hamiltonian and derive transformed evolution equations.

\subsection{Vortex dynamics equation}

The two-dimensional vorticity equation for incompressible or
quasi-geostrophic flows is
obtained using the bracket (\ref{eqn:breuler}) with the Hamiltonian \cite{shep90,holm-et-al}
\[
\H = \frac{1}{2} \iint \left( |\nabla \psi|^2 + R^{-2} \psi^2 \right) \dif x
\text{d}y.
\]
Here, $\psi$ is the streamfunction, related to the vorticity $\zeta$
through
\[
\zeta = \nabla^2 \psi - R^{-2} \psi,
\]
where the second term takes into account the effect of a free surface
in the quasi-geostrophic approximation, $R$ being the Rossby radius of
deformation. The evolution equation reads
\[
\zeta_t + [\psi,\zeta] = 0.
\]

A basic equilibrium consists of a parallel shear flow $U(y)$
superposed to the basic planetary rotation:
\[
\zeta_0 = f_0 + \beta y - U^\prime.
\]
Introducing the decomposition (\ref{eqn:decomp2}), one finds the
equation for the disturbance
\begin{equation} \label{eqn:eulerdist}
\zeta_t + U \zeta_x + \zetaop \psi_x + \ep [\psi,\zeta] = 0.
\end{equation}

Let us compute the free
energy (pseudoenergy) (\ref{eqn:dist2}) using the Casimirs functionals
which have the form
\[
\C = \iint C(\zeta_0 + \ep \zeta) \dif x \text{d}y,
\]
for any function $C(\cdot)$. It is well known that the proper choice of
$C(\cdot)$ is 
\begin{equation} \label{eqn:eulercas}
C(\cdot) = \int_{\zeta_0}^{(\cdot)} \psi_0(\mu) \dif \mu,
\end{equation}
where $\psi_0$ denotes the functional relationship between the
streamfunction and the vorticity which necessarily exists for
equilibrium solutions. The free energy is thus given by
\[
\F_\zeta = \iint \left\{ \left[\frac{1}{2} |\nabla \psi|^2 + R^{-2}
\psi^2 \right] + \ep^{-1} \int_0^\zeta
\left[ \psi_0(\zeta_0+\ep \mu)- \psi_0(\zeta_0) \right] \dif \mu
\right\} \dif x \text{d} y.
\]
As only the terms up to $O(\ep)$ are required for our purpose, we write
\begin{equation} \label{eqn:eulerfree}
\F_\zeta = \frac{1}{2} \iint \left[|\nabla \psi|^2 + R^{-2} \psi^2 -
\frac{U}{\zetaop} \zeta^2 - \frac{\ep}{3 \zetaop}
\frac{\text{d}}{\text{d}y} \left(\frac{U}{\zetaop} \right)\zeta^3 \right]
\dif x \text{d} y + O(\ep^2).
\end{equation}
We now introduce (\ref{eqn:CV4inv}) into
(\ref{eqn:eulerfree}). Because the relationship between $\psi$ and
$\eta$ is nonlinear, we define the two fields $\mphi$ and $\mtheta$ by
\begin{equation} \label{eqn:laplace}
\nabla^2 \mphi - R^{-2} \mphi = \eta \quad \mbox{ and } \quad 
\nabla^2 \mtheta
- R^{-2} \mtheta = \frac{1}{2} \dpar{}{y}\left(\frac{\eta^2}{2
\zetaop} \right),
\end{equation}
so that $\psi = \mphi + \ep \mtheta$ and thus
\[
|\nabla \psi|^2 + R^{-2} \psi^2 = |\nabla \mphi|^2 + R^{-2} \mphi^2 +
2 \ep \nabla \mphi \cdot \nabla 
\mtheta + 2 \ep R^{-2} \mphi \mtheta + O(\ep^2).
\]
The two other terms in (\ref{eqn:eulerfree}) require a little more
calculation (with some integrations by parts). Truncating to $O(\ep)$,
it is finally found that
\begin{equation} \label{eqn:eulerfree2}
\F_\eta = \frac{1}{2} \iint \left( |\nabla \mphi|^2 + R^{-2} \mphi^2 -
\frac{U}{\zetaop} \eta^2 + 2 \ep \nabla \mphi \cdot \nabla
\mtheta + 2 \ep R^{-2} \mphi \mtheta + \ep
\frac{U^\prime}{3{\zetaop}^2}  \eta^3  \right) \dif x
\text{d}y.
\end{equation}
Taking into account (\ref{eqn:laplace}), it is easy to calculate the
functional derivative of $\F_\eta$:
\[
\dfun{\F_\eta}{\eta} = - \mphi - \frac{U}{\zetaop} \eta - \ep
\mtheta + \ep \frac{1}{\zetaop} \mphi_y \eta + \ep
\frac{U^\prime}{2 {\zetaop}^2} \eta^2.
\]
The evolution equation for $\eta$ is then found using the Poisson 
operator (\ref{eqn:gardner}); it reads
\begin{equation} \label{eqn:evoleta}
\eta_t = - U \eta_x - \zetaop \mphi_x + \ep
\left(-\zetaop \mtheta_x + \mphi_{xy} \eta + \mphi_y \eta_x +
\frac{U^\prime}{\zetaop} \eta \eta_x \right).
\end{equation}
That this equation is a truncated version of (\ref{eqn:eulerdist}) 
can be verified by taking the time derivative of (\ref{eqn:CV4}),
using (\ref{eqn:eulerdist}) and (\ref{eqn:CV4inv}), and neglecting
terms of ordrer higher than $\ep$. Note that (\ref{eqn:evoleta}) is
relatively simple because it involves terms of order 1 and $\ep$ only:
no $O(\ep^2)$ are required to preserve the
Hamiltonian structure. Of course, if the equation were to be rewritten
in terms of $\zeta$, using (\ref{eqn:CV4inv}) as an exact
transformation, then the corresponding bracket and Hamiltonian would
include extra $O(\ep^2)$ terms. An advantage of a near-identity
transformation results from the fact that (\ref{eqn:eulerdist}) and
(\ref{eqn:evoleta}) 
have the same linear part; therefore the sum of results that have been
derived for the linearized vorticity equations can be exploited
directly in terms of $\eta$.

\subsection{Vlasov--Poisson equation} 
\label{ssec:VP}

The one-dimensional Vlasov--Poisson equation is also generated by the
bracket (\ref{eqn:breuler}). In standard notation, $y= v$ is the
particle velocity in the $x$ direction, and $\zeta = f(x,v)$ is the 
distribution function. For clarity, we will use the standard notation,
and we rewrite
\begin{equation} \label{eqn:vlbr}
\{\F,\G\} = \int \!\! \int f
\left[\dfun{\F}{f},\dfun{\G}{f}\right] \dif x \text{d}v,
\end{equation}
with
\[
[f,g] := \frac{1}{m} \left( \dpar{f}{x} \dpar{g}{v} - \dpar{f}{v}
\dpar{g}{x} \right).
\]
The Hamiltonian is given by
\begin{equation} \label{eqn:vlham}
\H = \frac{1}{2} \iint m v^2 f \dif x \text{d} v + \frac{1}{8 \pi}
\int E^2 \dif x,
\end{equation}
where the electric field $E(x)$ is related to the distribution function through the Poisson equation
\[
\dpar{E}{x} = - \dpar{^2\Phi}{x^2} = 4 \pi e  \left( \int f \dif v - N
\right), 
\]
$\Phi$ being the potential,  $N$ a fixed background density, and $m$ and $e$  the particle mass and charge, respectively.  

From (\ref{eqn:vlbr})--(\ref{eqn:vlham}), one can derive the Vlasov
equation
\[
f_t + v f_x + \frac{e}{m} E f_v = 0.
\]

Considering the decomposition (\ref{eqn:decomp2}) in a basic
distribution $f_0(v)$ and an $O(\ep)$ disturbance, we write the
disturbance evolution equation
\begin{equation} \label{eqn:vldist}
f_t + v f_x + \frac{e}{m} f_0^\prime E + \ep \frac{e}{m} E f_v = 0.   
\end{equation} 
The free energy can be constructed similarly to that for the vorticity
equation, with $\psi_0(\cdot)$ in (\ref{eqn:eulercas}) replaced by
$\frac{1}{2} m v^2 + e \Phi_0$ which is functionally dependent on
$f_0$. It is then found that
\begin{eqnarray} \label{eqn:vlham2}
\F_f &=& \frac{1}{8 \pi} \int E^2 \dif x - \iint \left[ \frac{1}{2}
	\dpar{}{f_0} 
	\left(\frac{mv^2}{2}\right)  f^2 +
	\frac{\ep}{6} \dpar{^2}{f_0^2}  \left(\frac{mv^2}{2}\right)
	f^3 \right] 
 	\dif x \text{d}v +
            O(\ep^2) \nonumber \\
     &=& \frac{1}{8 \pi} \int E^2 \dif x - m \iint \left[\frac{v}{2
	f_0^\prime} f^2 + 
	\frac{\ep}{6 {f_0^\prime}^2} \left(1-\frac{v f_0^{\prime
	\prime}}{f_0^\prime} \right) f^3 \right] \dif x \text{d}v +
            O(\ep^2).
\end{eqnarray}

We now introduce the transformation (\ref{eqn:CV4})--(\ref{eqn:CV4inv})
which can be rewritten as
\begin{equation} \label{eqn:vlCV}
\eta = f - \ep \dpar{}{v} \left( \frac{f^2}{2 f_0^\prime} \right)
\quad \mbox{and} \quad 
f = \eta + \ep \dpar{}{v} \left( \frac{\eta^2}{2 f_0^\prime} \right) +
O(\ep^2).
\end{equation}
It is such that the dynamics of $\eta$ is generated up to $O(\ep)$ by
the Poisson  operator
\begin{equation} \label{eqn:vlJ}
\J_0 = \frac{f_0^\prime}{m} \dpar{}{x}.
\end{equation}
Introducing (\ref{eqn:vlCV}) in (\ref{eqn:vlham2}), we can compute the Hamiltonian for
$\eta$. Compared to the vortex dynamics case, a simplification occurs
because the Poisson equation is unchanged by the transformation:
\[
\dpar{E}{x} = - \dpar{^2\Phi}{x^2} = 4 \pi e  \left( \int f \dif v 
\right)= 4 \pi e  \left( \int \eta \dif v 
\right),
\]
so that the first term in the right-hand side of (\ref{eqn:vlham2})
does not lead to a $O(\ep)$ contribution. The other two terms are
readily simplified using integration by parts, and finally yield the
truncated Hamiltonian
\begin{equation} \label{eqn:vlham3}
\F_\eta = \frac{1}{8 \pi} \int E^2 \dif x - \frac{m}{2} \iint
\left(\frac{v}{f_0^\prime} \eta^2 - \frac{\ep}{3 {f_0^\prime}^2}
\eta^3 \right) \dif x \text{d}v.
\end{equation}
The evolution equation for $\eta$ is then readily obtained from
(\ref{eqn:vlJ}) and (\ref{eqn:vlham3}). It is given by
\begin{equation} \label{eqn:vleq}
\eta_t = -v \eta_x - \frac{e}{m} f_0^\prime E + \frac{\ep}{f_0^\prime}
\eta \eta_x.
\end{equation}
Of course, this equation can also be found from (\ref{eqn:vldist}) and
(\ref{eqn:vlCV}). 

\section{Stratified fluid and two-dimensional MHD} \label{sec:kir}

The bracket for stratified fluid and two-dimensional MHD \cite{MH84,benj86}, defined by
\begin{equation} \label{eqn:kibr}
\{\F,\G\}=\iint \left(
	\zeta \left[\dfun{\F}{\zeta},\dfun{\G}{\zeta}\right]
	+\rho \left[\dfun{\F}{\rho},\dfun{\G}{\zeta}\right]
       +\rho \left[\dfun{\F}{\zeta},\dfun{\G}{\rho}\right] 
		\right) \dif x \text{d}y,
\end{equation}
corresponds to the Poisson  operator
\[
\J = \left(
	\begin{array}{c c}
   -[\zeta,\cdot]       &  	 -[\rho,\cdot] \\
   -[\rho,\cdot]	&       	0
	\end{array}
     \right),
\]
where $(\zeta,\rho)$ are the dyamical variables.
As for the Euler bracket, we consider a basic solution that depends
only on $y$, $\zeta_0(y)$ and $\rho_0(y)$, and we introduce a
decomposition of the form (\ref{eqn:decomp2}). From now on
$(\zeta,\rho)$ thus designates the $O(\ep)$ disturbance to the basic
solution. The leading-order Poisson operator which generates the
linear evolution for $(\zeta,\rho)$ is given by
\begin{equation} \label{eqn:kiJ}
\J_0 = \left(
	\begin{array}{c c}
   \zetaop       &  	\rhoop \\
   \rhoop	&	 0 \\  
	\end{array}
     \right) \dpar{}{x},
\end{equation}
We are seeking new variables $(\eta,\si)$ such that condition
(\ref{eqn:cond2}) is satisfied. Rather than using a general
near-identity transformation, we can exploit the close analogy between
the bracket (\ref{eqn:kibr}) and the Euler bracket, and use the
procedure based on 
dynamical accessibility. From (\ref{eqn:kiJ}), it is clear that
the new variables will have a constant average in the $x$ direction
and we can expect the new variables to be given by the old ones minus
their average part (up to a $\partial_x$ term which can be shown to be
irrelevant), as observed in Section \ref{sec:eul}. To obtain this average, we consider the dynamical
accessibility condition
\begin{eqnarray} \label{eqn:kida}
\ep \left(
	\begin{array}{c}
		\zeta \\ \rho
	\end{array} \right)
&=& \left( \exp \{-\G,\cdot\} - \mathsf{Id} \right)
      \left
	(\begin{array}{c}
		\zeta_0 \\ \rho_0
	\end{array} \right) \nonumber \\
&=& \ep \left(
	\begin{array}{c}
		g_x \zetaop + h_x \rhoop \\ g_x \rhoop
	\end{array} \right)
+ \frac{\ep^2}{2} \left(
	\begin{array}{c}
		\left[g,g_x \zetaop\right] + \left[h,g_x \rhoop\right]
		+ \left[g,h_x \rhoop\right]  \\
		\left[g,g_x \rhoop\right]
	\end{array} \right)
+ O(\ep^3),
\end{eqnarray}
where $g$ and $h$ are arbitrary functions and
\[
\G = \iint \left( g \zeta + h \rho\right) \dif x \text{d} y.
\]
Taking the $x$-average of (\ref{eqn:kida}), one finds with some effort
\[
\left( \begin{array}{c}
		\overline{\zeta} \\ \overline{\rho} 
	\end{array} \right)
= \frac{\ep}{2} \dpar{}{y} \left(
\begin{array}{c}
	\zetaop \overline{g_x^2} + 2 \rhoop \overline{g_x h_x} \\ \rhoop \overline{g_x^2}
\end{array} \right) + O(\ep^2).
\]
Eq.\ (\ref{eqn:kida}) also yields the relationship between $(g,h)$ and
$(\zeta,\rho)$, namely
\[
\left(
	\begin{array}{c} g_x \\ h_x \end{array}
\right) 
= \frac{1}{\rhoop}
\left(
	\begin{array}{c} \rho \\ \zeta - \zetaop \rho /\rhoop
	\end{array}
\right) + O(\ep),
\]		
so that the average fields can be written
\[
\left( \begin{array}{c}
		\overline{\zeta} \\ \overline{\rho} 
	\end{array} \right)
= \ep \dpar{}{y} \left[\frac{1}{2\rhoop} \left(
\begin{array}{c}
	2 \overline{\rho \zeta} - \zetaop \overline{\rho^2}/\rhoop \\
	\overline{\rho^2} 
\end{array} \right) \right]+ O(\ep^2).
\]

This suggests the transformation from $(\zeta,\rho)$ to the new
variables $(\eta,\si)$ defined by
\begin{equation} \label{eqn:kiCV}
\left( \begin{array}{c}
		\eta \\ \si 
	\end{array} \right)
= \left( \begin{array}{c}
		\zeta \\ \rho 
	\end{array} \right)
-
\ep \dpar{}{y} \left[\frac{1}{2 \rhoop} \left(
\begin{array}{c}
	2 \rho \zeta - \zetaop \rho^2/\rhoop \\ \rho^2
\end{array} \right) \right].
\end{equation}
It can be verified that the bracket for those variables is given by
$\J_0 + O(\ep^2)$, with $\J_0$ defined by (\ref{eqn:kiJ}), as required.
The inverse transformation is readily obtained as
\begin{equation} \label{eqn:kiCVinv}
\left( \begin{array}{c}
		\zeta \\ \rho 
	\end{array} \right)
= \left( \begin{array}{c}
		\eta \\ \si 
	\end{array} \right)
+
\ep \dpar{}{y} \left[\frac{1}{2 \rhoop} \left(
\begin{array}{c}
	2 \si \eta - \zetaop \si^2/\rhoop \\ \si^2
\end{array} \right) \right] + O(\ep^2).
\end{equation}
With these results, we can derive the Hamiltonian equations describing
the weakly nonlinear evolution of a disturbance in a stratified fluid
and in two-dimensional MHD.

\subsection{Stratified fluid}

The equations for an incompressible stratified fluid are
derived from the bracket (\ref{eqn:kibr}) and the Hamiltonian \cite{holm-et-al, benj86}
\begin{equation} \label{eqn:sfham}
\H = \iint \left( \frac{1}{2} \rho |\nabla \psi|^2 + \rho g y \right)
\dif x \mathrm{d} y.
\end{equation}
The dynamical variables $\zeta$ and $\rho$ represent the
(mass-weighted) vorticity and
density, and the coordinate $y$ is vertical. The streamfunction $\psi$
is related to the vorticity through
\[
\zeta = \nabla \cdot (\rho \nabla \psi).
\]
The evolution equations take the form
\begin{eqnarray} \label{eqn:sfevol}
\zeta_t + [\psi,\zeta] + [\rho,g y-   |\nabla
\psi|^2/2] &=& 0 \\
\rho_t + [\psi,\rho] &=& 0.
\end{eqnarray}
We now consider a basic solution consisting of the  parallel flow
$U(y)$ and stratification $\rho_0(y)$, so that the basic vorticity is
given by
\[
\zeta_0 = - (\rho_0 U)^\prime.
\]
To $O(\ep)$, a disturbance to this basic solution obeys the evolution equations 
\begin{eqnarray} \label{eqn:sfde}
\zeta_t + U \zeta_x + \zetaop \psi_x + (g - U U^\prime) \rho_x -
\rhoop U \psi_{xy} && \nonumber \\ 
+ \ep \left( [\psi,\zeta] + \rhoop \nabla \psi
\cdot \nabla  \psi_x + [\rho,U \psi_y] \right) &=& 0 \nonumber \\
\rho_t + U \rho_x + \rhoop \psi_x + \ep [\psi,\rho] &=& 0,
\end{eqnarray}
with the streamfuction/vorticity relation given by
\begin{equation} \label{eqn:sfsf}
\zeta = \nabla \cdot (\rho_0 \nabla \psi) - (U \rho)_y + \ep
\nabla \cdot (\rho \nabla \psi).
\end{equation}

The free energy can now be computed using the Casimir functionals. 
Computations detailed in Appendix \ref{app:strat} give
\begin{eqnarray} \label{eqn:sffreefinal}
\F_{(\eta,\si)} = && \iint \Bigg\{ \frac{1}{2} \rho_0 |\nabla \mphi|^2 -
\frac{U}{\rhoop} \eta \si + \frac{1}{2 \rhoop}  \left( U U^\prime +
\frac{\zetaop U}{\rhoop} - g \right) \si^2  \\
&&  +   \ep \left[ \frac{1}{2}  \si |\nabla \mphi|^2 + \rho_0
\nabla \mphi \cdot \nabla \mtheta + \frac{U^\prime}{2 {\rhoop}^2} \eta
\si^2 - \frac{1}{6 {\rhoop}^3 } \left[ 2 \zetaop U^\prime - \rhoop (U
U^\prime)^\prime \right] \si^3 \right]
\Bigg\} \dif x \mathrm{d}y. \nonumber
\end{eqnarray}
Here, $\mphi$ and $\mtheta$ are defined by the decomposition
$\psi = \mphi + \ep \mtheta$ of the streamfunction.
From (\ref{eqn:sfsf}) and (\ref{eqn:kiCV}), we find these
to be given in terms of the new dependent
variables $(\eta,\si)$ by
\begin{eqnarray} \label{eqn:sfphi}
\nabla \cdot (\rho_0 \nabla \mphi) &=& \eta + \dpar{}{y}(U \si)
\nonumber \\
\nabla \cdot (\rho_0 \nabla \mtheta) &=& \dpar{}{y}\left( \frac{\eta
\si}{\rhoop}-\frac{\zetaop \si^2}{2 \rhoop} \right) + \dpar{}{y}
\left[U \dpar{}{y} \left( \frac{\si^2}{2 \rhoop} \right) \right] -
\nabla \cdot (\si 
\nabla \mphi).
\end{eqnarray}

The equations of motion in terms of $(\eta,\si)$ are derived by
applying the Poisson operator  $\J_0$ given by (\ref{eqn:kiJ}) to the functional dervatives of $\F_{(\eta,\si)}$, computed in Appendix \ref{app:strat}.
The final result reads
\begin{eqnarray} \label{eqn:sfeqfinal}
\eta_t &=& - U \eta_x - \zetaop \mphi_x - (g - U U^\prime) \si_x +
\rhoop U \mphi_{xy} \nonumber \\
&& + \ep \bigg\{ -\zetaop \mtheta_x - \rhoop U \mtheta_{xy}  - \rhoop
\nabla \mphi \cdot \nabla \mphi_x + (\mphi_y \eta)_x  \nonumber
\\
&& - \left. \left[ (U \mphi_y)_y \si \right]_x +
\frac{U^\prime}{\rhoop} (\eta \si)_x - \frac{1}{{\rhoop}^2} \left[
\zetaop U^\prime + \rhoop (U U^\prime)^\prime \right] \si \si_x
\right\}  \\
\si_t &=& - U \si_x - \rhoop \mphi_x + \ep \left[ - \rhoop \mtheta_x +
(\mphi_y \si)_x + \frac{U^\prime}{\rhoop} \si \si_x \right]. \nonumber
\end{eqnarray}
It can be checked that these equations are also found directly by
mnaipulating (\ref{eqn:kiCV}), (\ref{eqn:kiCVinv}) and (\ref{eqn:sfde}).
 
\subsection{Two-dimensional MHD}

The equations for two-dimensional MHD with the magnetic field in the
plane are Hamiltonian with the bracket (\ref{eqn:kibr}), where $\zeta$
is to be interpreted as the vorticity and $\rho$ as the magnetic
potential (often denoted by $A$). The corresponding Hamiltonian \cite{MH84,MM84} is
\begin{equation} \label{eqn:mhdham}
\H = \frac{1}{2}\iint \left( |\nabla \psi|^2 + |\nabla \rho|^2 \right)
\dif x \mathrm{d}y,
\end{equation}
where $\psi$ is the streamfunction which, according to our
previous sign convention, is defined by $\nabla^2 \psi =  \zeta$. 
The evolution equations are then of  the form
\begin{eqnarray*}
\zeta_t + [\psi,\zeta] + [\rho, \nabla^2 \rho] &=& 0 \\
\rho_t +  [\psi,\rho] &=& 0.
\end{eqnarray*}  
The basic solution we consider is again a parallel flow
$U(y)$, with the magnetic potential $\rho_0(y)$, corresponding to a magnetic
field parallel to the streamlines. The evolution of a small
disturbance to  this basic solution is governed by
\begin{eqnarray} \label{eqn:mhdevol}
\zeta_t + U \zeta_x + \zetaop \psi_x  - {\rho_0}''' \rho_x +
\rhoop  \nabla^2 \rho_x  + \ep \left\{ [\psi, \zeta] + [\nabla^2 \rho,
\rho] \right\} &=& 0 \nonumber \\
\rho_t + U \rho_x + \rhoop \psi_x  + \ep [\psi,\rho] &=& 0\,, 
\end{eqnarray} 
where each prime denotes a derivative with respect to $y$. 

The derivation of the free energy for this system closely follows that
for the stratified fluid: decomposing $\psi = \mphi + \ep \mtheta$, with
\[
\nabla^2 \mphi = \eta, \quad \nabla^2 \mtheta = \dpar{}{y}
\left(\frac{\eta \si}{\rhoop} - \frac{\zetaop \si^2}{2 {\rhoop}^2}\right),
\]
computations detailed in Appendix \ref{app:mhd} lead to
\begin{eqnarray}  \label{eqn:mhdfree2}
\F_{(\eta,\si}) &=& \iint \left\{ 
\frac{1}{2} \left(|\nabla \phi|^2 + |\nabla \si|^2 \right) -
\frac{U}{\rhoop} \eta \si + \frac{1}{2\rhoop} \left(\frac{\zetaop
U}{\rhoop} + \rho_0^{'''} \right) \si^2 \right.  \\
&& + \left. \ep \left[ \nabla \mphi \cdot \nabla \mtheta +
\frac{U^\prime}{2 {\rhoop}^2} \eta \si^2 + \frac{1}{2 \rhoop} \si^2
\nabla^2 \si_y - \frac{1}{6 {\rhoop}^3} \left( 2 \zetaop U^\prime +
\rhoop {\rho_0}^{''''} \right) \si^3 \right] \right\} \dif x \mathrm{d} y\,.
\nonumber
\end{eqnarray}

Applying the Poisson operator (\ref{eqn:kiJ}), one can finally
find the evolution equations
\begin{eqnarray} \label{eqn:mhdeqfinal}
\eta_t &=& - U \eta_x - \zetaop \mphi_x - \rhoop \nabla^2 \si_x +
\rho_0^{''''} \si_x \nonumber \\
&& + \ep \left[ - \zetaop \mtheta_x + (\mphi_y \eta)_x +
\frac{U^\prime}{\rhoop}(\eta \si)_x - \frac{1}{{\rhoop}^2}
\left(\zetaop U^\prime + \rhoop \rho_0^{''''} \right) \si \si_x
\right. \nonumber \\
&& + \left. (\si \nabla^2 \si_y)_x - \rhoop \nabla^2 \frac{^2}{x
\partial y} \left(\frac{\si^2}{2 \rhoop}\right) \right] \\
\si_t &=& -U \si_x - \rhoop \mphi_x + \ep \left[-\rhoop \mtheta_x +
(\mphi_y \si)_x + \frac{U^\prime}{\rhoop} \si \si_x \right]. \nonumber
\end{eqnarray}
Again, these equations can be verified by using directly
(\ref{eqn:kiCV}), (\ref{eqn:kiCVinv}) and (\ref{eqn:mhdevol}).

\section{Discussion}
\label{sec:discussion}

We have described beatification,  a general method for obtaining weakly nonlinear Hamiltonian dynamical systems from a variety of finite and infinite-dimensional noncanonical Hamiltonian systems.  We have seen that unlike canonical systems,  for which the Poisson matrix is constant, perturbation theory for noncanoncial systems possesses the  additional necessity of simultaneously expanding the Poisson bracket and the Hamiltonian.  This could have been done for any known orbit or point in phase space, but here we expanded about a given  equilibrium solution and obtained weakly nonlinear systems for our examples.   Table \ref{ta:1} gives the locations of the results for the  four  infinite-dimensional systems we considered.  
 
\begin{table}[tb] \label{ta:1}
\begin{center}
\begin{tabular}{l c c c c}
\hline
	& Vortex  dynamics & Vlasov--Poisson & Stratified fluid & 2-D MHD
	\\
\hline \hline
Bracket & (\ref{eqn:gardner}) & (\ref{eqn:vlJ}) & (\ref{eqn:kiJ}) &
	(\ref{eqn:kiJ}) \\
Free energy & (\ref{eqn:eulerfree2}) & (\ref{eqn:vlham3}) &
	(\ref{eqn:sffreefinal}) & (\ref{eqn:mhdfree2}) \\
Equations & (\ref{eqn:evoleta}) & (\ref{eqn:vleq}) &
	(\ref{eqn:sfeqfinal}) & (\ref{eqn:mhdeqfinal}) \\
\hline
\end{tabular}
\end{center}
\caption{Equation numbers for the
Hamiltonian structures and evolution 
equations in terms of the transformed variables.} 
\end{table}

Beatification can also be applied  to the  large  class of Hamiltonian plasma reduced fluid models that have  Poisson bracket (Jacobian) nonlinearities.  Such models have  additional   scalar fields that describe two-fluid or extended MHD effects,  the effects of gyroviscosity, which gives rise to nondissipative momentum transport in magnetized plasma \cite{MCT85,HHM87,SPK94,KPS94,GCPP94,KK04,WMH04,TMWG08,KWM15}, and electromagnetic effects  \cite{WHM09,WT12}.  All these models fit within  the  theory of Ref.~\cite{TM00}, and the general beatification transformation can be worked out for all of them.   Although these models are  two-dimensional,  they can be extended by  geometric aspect ratio expansion to what are referred to as  weakly three-dimensional models \cite{TMGP10}, where the third dimension enters the equations of motion  linearly. 

More general Hamiltonian field theories have more complicated Poisson brackets than those of Table \ref{ta:1} considered here.  For example, the bracket for three-dimensional MHD  \cite{MG80} has Lie-Poisson form, but  in full generality it  has in addition to the velocity and magnetic fields the density and entropy,   giving a total of eight variables.  Although this complicates significantly the beatification transformation, calculations like those presented in the present work can be done for this system.  In fact, even the rather complicated magnetofluid model of full extended MHD, which  includes Hall drift as well as electron inertia effects,  is amenable, since we now have the noncanonical Poisson bracket,  one that is not of Lie-Poisson form, for this full system \cite{HKY14,LMM15}.    This full system  includes Hall MHD \cite{YH13} and inertial MHD as Hamiltonian subsystems and,  because of the transformations given in  \cite{LMM15},  we see that beatification of only the Hall bracket is sufficient for beatification of inertial MHD and, indeed, the entire extended MHD.  Similarly, the  complicated  bracket  for the    Vlasov-Maxwell system \cite{M80,mor82,MW82,M13} can be beatified.   

Another application  of the  beatification procedure is to Hamiltonian systems governed by brackets that emerge from Dirac's constraint theory.  Such systems have Poisson brackets called Dirac brackets, an examples being the Poisson brackets for  the incompressible fluid  \cite{NT99,CMT12,CGBTM13}, incompressible MHD \cite{CMT12,CGBTM13}, and the modified Hasegawa-Mima equation \cite{CMT14}.  Brackets, even more general than the conventional Dirac bracket  \cite{C13} can be beatified.

 The beatification procedure proposed in this paper lays the groundwork for the  canonization of weakly nonlinear Hamiltonian systems. Because the brackets $J_0$ in the new coordinates are exactly those governing the linearized dynamics, the further coordinate transformations required to canonize weakly nonlinear systems are exactly those that canonize linear systems. We demonstrate this by  explicitly carrying out the full canonization of the Vlasov-Poisson system considered in Section \ref{ssec:VP} in terms of action-angle variables, setting the stage for the perturbation theory proposed in  \cite{M00}. This is detailed in Appendix  \ref{sec:canon}.    Details of this are contained in  \cite{Y01}, where equations for interacting continuous spectra analogous to  the  three-wave problem in finite dimensions were obtained.  Similar calculations can be performed for fluid shear flow dynamics that have similar continuous spectra \cite{BM02,v96,v98}, and the relationship of this approach to previous echo calculations \cite{VMW98} and recent rigorous results \cite{B13}  would be of interest to pursue.

The degree to which the weakly nonlinear systems obtained here accurately track the full dynamics  will depend on a case-by-case basis on the system under consideration, but some generic comments can be made.  Clearly one would expect a better approximation the closer one chooses an initial condition to the equilibrium state and the smaller the numerical coefficients of the nonlinear terms in the Hamiltonian.  For stable equilibria with Hamiltonians that have a convex linear part,  our systems should give accurate  frequency shifts due to nonlinearity for near equilibrium dynamics.  However, to capture the extent of a basin of stability, if finite amplitude instability exists, would require extending the beatification transformation to higher order.    For expansion about stable equilibria 
with Hamiltonians that have a nonconvex linear part, i.e., systems  with negative energy modes, the situation is more delicate, even for the case of finite dimensions \cite{WW77,KM95} where explosive growth can occur.  Progress has been made in understanding the structural stability of infinite-dimensional systems  \cite{HM10},  but nonlinear behavior with negative energy modes complicates matters.  For unstable equilibria, accurate slowing of the growth rate may be possible, but to understand the extent to which a mode can grow would also require a higher order expansion. 

The beatification transformations considered in this paper flattened the Poisson bracket to first order.  In recent work this has been generalized by flattening to second order  \cite{VCM15}, which is necessary  to obtain consistent dynamics if one expands about a state that is not an equilibrium state.   Also, this higher order beatification can be used to address the accuracy problems mentioned above.  Moreover,   it has been shown  \cite{VCM15b} how to construct a beatification transformation in terms of an operator series that flattens the Poisson bracket to all orders.  This and other examples will be recorded and explored in future work. 

\section*{Acknowledgments}

One of us (PJM) was supported  by the US Department of Energy Contract No.~DE-FG02-04ER54742.   Both of us would like to acknowledge the hospitality of the Isaac Newton Institute for Mathematical Sciences, where the bulk of this work was performed during the 1996 program {\it The Mathematics of Atmosphere and Ocean Dynamics}.

\section*{Appendices}
\appendix    

\section{Two-spin system} 
\label{app:twospin}

In this Appendix, we provide some details of the treatment of the two-spin system considered in Section \ref{sec:twospin}. 

\subsection{Canonical variables for the linearised system}

Consider the linearized equations for $\za$. They are
 generated by
the Hamiltonian (\ref{eqn:hz}) with the bracket 
\begin{equation} \label{eqn:pblin}
\{f,g\}_0 = m B_1 \sum_{\al=1}^{2} \eps_{ij1}  \dpar{f}{\za^i} \dpar{g}{\za^j}
\end{equation}
obtained by linearising (\ref{eqn:pbnlin}). 
One finds explicitly
\[
\left\{ \begin{array}{l}
\dot{z}_{(\al)}^1 = 0 \\
\dot{z}_{(\al)}^2 = m B_1 \left(\la \za^3 + a z_{(\al + 1)}^3 \right) \\
\dot{z}_{(\al)}^3 = -m B_1 \left(\la \za^2 + a z_{(\al + 1)}^2
\right)
\end{array} \right.
 \quad \al=1,2.
\]
The invariance of $\za^1$ follows from the invariance of the Casimir
functions in the original system. We now seek action--angle variables
for this system. 
Using $z := (\z^2,\z^3,\Z^2,\Z^3)^{\mathrm{T}}$, we can rewrite the dynamics as
\[
\dot{z} = \Am z,
\]
where
\[
\Am := mB_1 \left(
\begin{array}{c c c c}
0    & \la &   0 & a \\
-\la &  0  &  -a & 0 \\
0    &  a  &   0 & \la \\
-a   &  0  &- \la& 0 
\end{array}
\right).
\]
The vector $z$ can then be expanded in terms of the eigenvectors of
$\Am$:
\[
z = \La_{(1)} x_{(1)} + \La_{(2)} x_{(2)} + \mbox{ c.c.},
\]
where
\[
x_{(1)}:=(-\ic,-1,\ic,1)^{\mathrm{T}}/2 \quad \mbox{and} \quad 
x_{(2)}:=(\ic,1,\ic,1)^{\mathrm{T}}/2
\]
correspond to the eigenvalues $\ic \si_{(1)} := \ic B_1 (1+2 am)$ and
$\ic \si_{(2)} := \ic B_1$, respectively. In terms of the variables
$\La_{(\al)}$, the Poisson bracket (\ref{eqn:pblin}) is
\[
\{f,g\}_0 = - \ic m B_1 \sum_{\al=1}^2 \left( \dpar{f}{\Lama}
\dpar{g}{\Lama^*} - \dpar{f}{\Lama^*}
\dpar{g}{\Lama} \right).
\]
Action and angle variables $(\Ja,\tha)$ are then defined by
\begin{equation} \label{eqn:aa}
\Lama = \sqrt{|mB_1| \Ja} \exp (-\ic s \tha),
\end{equation}
where $s := \mbox{sign}(m B_1)$, so that the bracket becomes canonical:
\begin{equation} \label{eqn:pbaa}
\{f,g\}_0 = \sum_{\al=1}^2 \left( \dpar{f}{\tha} \dpar{g}{\Ja} -
\dpar{f}{\Ja} \dpar{g}{\tha} \right).
\end{equation}
Correspondingly, the (quadratic) Hamiltonian (\ref{eqn:pbnlin}) reduces to
\[
F = - s  \sum_{\al=1}^2 \si_{(\al)} \Ja + a \z^1 \Z^1.
\]
The last term is of course irrelevant, as the variables $\z^1$ and
$\Z^1$ do not appear in the bracket.

\subsection{Canonical perturbation theory for the weakly nonlinear system}

Using the mixed generating function
$S(\tha,I_{(\alpha)})$, we transform $(\Ja,\tha)$ into the new variables
$(\Ia,\pha)$, with
\begin{equation} \label{eqn:no}
\dpar{S}{\tha} = \Ja \quad \mbox{ and } \quad \dpar{S}{\Ia}=\pha,
\end{equation}
such that the new Hamiltonian is independent of $\pha$. It is easy to
check that this is achieved with
\begin{equation} \label{eqn:gf}
S = \sum_{\al=1}^{2} \tha I_{(\alpha)} - \frac{\ep}{2mB_1} (c_1-c_2)
\sqrt{\smash[b]{I_{(1)} I_{(2)}}} \sin(\theta_{(1)}-\theta_{(2)}).
\end{equation}
The new Hamiltonian is obtained by averaging; it is given by
\[
F(\Ia) = - s  \sum_{\al=1}^2 \si_{(\al)} \Ia + a c_1 c_2 - \ep s a
(c_1 + c_2) I_{(1)}.
\]
The frequencies of the nonlinear system are thus given by $\si_{(1)} + \ep a
(c_1+c_2)$ and $\si_{(2)}$. The exact relationship between
$(\Ia,\pha)$ and $(\Ja,\tha)$ can be found from
(\ref{eqn:no})--(\ref{eqn:gf}). 

It is interesting to note that,  due to the extreme simplicity of the
system, one can integrate the weakly nonlinear equations without
resorting to the canonical perturbation techniques. From
(\ref{eqn:titi}),
it is clear that the nonlinear coupling vanishes to the order of
accuracy considered here if $\etaa^1 = O(\ep)$
initially. Actually, this can be obtained for any initial condition by
suitably redefining the decomposition between equilibrium solution and
the disturbance. This decomposition is indeed partly arbitrary, and
one can write the initial condition as
\[
\oma^1(0) = mB_1 + \ep \za^1(0) = m^\prime_{(\al)} B_1,
\]
with ${m}^\prime_{(\al)} := 1 + \ep \za^1(0)/(mB_1)$. With the second
decomposition the disturbance has initially vanishing components in
the direction of the external magnetic field, and $\etaa^1 = O(\ep)$
for all time. The cubic terms in the Hamiltonian vanish, so that there
is no weakly nonlinear effect at the order considered. Studying the
linear system as above but with $m^\prime_{(1)} \not=
m^\prime_{(2)}$, one finds the two frequencies $\si_{(1)} + \ep a
(c_1+c_2)$ and $\si_{(2)}$ as obtained by the perturbation theory.

\section{Extension of the Zakharov--Piterbarg transformation} \label{app:zp}

The variables $x,y,\zeta(x,y)$ are transformed into $x,z,\etac(x,z)$ defined by
\begin{equation} \label{eqn:atr}
z[x,y,\zeta] = y_0[\zeta_0(y) + \ep \zeta(x,y)], \quad
\etac[x,y,\zeta] =\ep^{-1} 
\left.\dt{\zeta_0}{y_0}\right|_z (z - y).
\end{equation}
These relations can be inverted according to
\begin{equation} \label{eqn:atri}
y[x,z,\etac]=z-\ep \left.\dt{y_0}{\zeta_0}\right|_{\zeta_0(z)} \etac(x,z),
\quad \zeta[x,z,\etac]=\ep^{-1} \left[ \zeta_0(z) - \zeta_0(y)\right].
\end{equation}
Roughly  speaking, one can say that the transformation exchanges the
roles of the independent and dependent variables: $z$ is  essentially
defined in 
terms of the vorticity, while $\eta_c$ is defined in terms of parcel
displacement --- the expressions are somewhat complicated by the fact
that the transformation is near-identity so that  $z$ is a distance
and $\etac$ a disturbance vorticity. Noting that
\begin{equation} \label{eqn:aj}
\mathrm{d}y = \left[1 - \ep
\dpar{}{z}\left(\left.\dt{y_0}{\zeta_0}\right|_{\zeta_0(z)} \etac \right)
\right] \mathrm{d}z,
\end{equation}
we find a condition analogous to (\ref{eqn:wnl}) for the
 well-definiteness of the transformation. 
 We now compute the bracket
(\ref{eqn:breuler}) in terms of $x,z,\etac$. From
(\ref{eqn:atr})--(\ref{eqn:atri}), one can see that 
$\zeta(x,y)$ and $\etac(x,z)$ are functionally related through 
\[
\zeta(x,y) =  \ep^{-1} \int \left\{ \zeta_0(z) -  \zeta_0\left[z - \ep
\left. \dt{y_0}{\zeta_0}\right|_{\zeta_0(z)} \etac(x,z) \right]
\right\} \de(z - y_0[\zeta_0(y) + \ep \zeta(x,y)]) \dif z.
\]
Taking the variation of this equality leads to
\begin{eqnarray*}
\de \zeta(x,y) &=& \int \left. \dt{\zeta_0}{y_0} \right|_{y}
\left. \dt{y_0}{\zeta_0}\right|_{\zeta_0(z)} \de \eta(x,z) \, \de(z -
y_0[\zeta_0(y) + \ep \zeta(x,y)]) \dif z \\
	       &+& \left. \dpar{}{z} \left\{ \zeta_0(z) - 
\zeta_0\left[z - \ep 
\left. \dt{y_0}{\zeta_0}\right|_{\zeta_0(z)} \etac(x,z) \right]
\right\}
\right|_{z=y_0[\zeta_0(y) + \ep \zeta(x,y)]} \left.\dt{y_0}{\zeta_0}
\right|_{\zeta_0(y) + \ep \zeta(x,y)} \de \zeta(x,y),
\end{eqnarray*} 
which can be rewritten as
\begin{eqnarray*}
&& \left. \dpar{\zeta_0}{y_0} \right|_y   \left.\dt{y_0}{\zeta_0}
\right|_{\zeta_0(y) + \ep \zeta(x,y)}  \left[1 - \ep
\dpar{}{z}\left(\left.\dt{y_0}{\zeta_0}\right|_{\zeta_0(z)} \etac(x,z) \right)
\right]_{z=y_0[\zeta_0(y) + \ep \zeta(x,y)]} \, \de \zeta(x,y) \\
&=& \int \left. \dt{\zeta_0}{y_0} \right|_{y}
\left. \dt{y_0}{\zeta_0}\right|_{\zeta_0(z)} \de \eta(x,z) \, \de(z -
y_0[\zeta_0(y) + \ep \zeta(x,y)]) \dif z,
\end{eqnarray*} 
and finally as
\begin{equation} \label{eqn:avar}
\de \zeta(x,y) = \int \left[1 - \ep
\dpar{}{z}\left(\left.\dt{y_0}{\zeta_0}\right|_{\zeta_0(z)} \etac(x,z)
\right)\right]^{-1} \,  \de \eta(x,z) \, \de(z -
y_0[\zeta_0(y) + \ep \zeta(x,y)]) \dif z.
\end{equation}
Consider now the variation of an arbitrary functional $\F$, namely
\begin{equation} \label{eqn:afun}
\de \F = \iint \dfun{\F}{\zeta(x,y)}\, \de\zeta(x,y)  \dif x \mathrm{d}y
=  \iint \dfun{\F}{\etac(x,z)}\, \de\etac(x,z)  \dif x \mathrm{d}z.
\end{equation}
Introducing (\ref{eqn:avar}), the first equality becomes
\[
\de \F = \iiint \left[1 - \ep
\dpar{}{z}\left(\left.\dt{y_0}{\zeta_0}\right|_{\zeta_0(z)} \etac(x,z)
\right)\right]^{-1} \,  \de \eta(x,z) \, \de(z -
y_0[\zeta_0(y) + \ep \zeta(x,y)]) \dfun{\F}{\zeta(x,y)}  \dif x
\mathrm{d}y \mathrm{d}z.
\]
From (\ref{eqn:aj}), it can be seen that the integration in $y$ can be
carried out: defining $z^\prime$ by $z^\prime = y_0[\zeta_0(y) + \ep
\zeta(x,y)]$, one finds
\[
\int \de(z -
y_0[\zeta_0(y) + \ep \zeta(x,y)]) \dfun{\F}{\zeta(x,y)} \dif y = \int \left[1 - \ep
\dpar{}{z^\prime}\left(\left.\dt{y_0}{\zeta_0}\right|_{\zeta_0(z^\prime)} \etac(x,z^\prime)
\right)\right] \de(z-z^\prime) \dfun{\F}{\zeta(x,y)} \dif z^\prime.
\]
Substituting this result in the previous equation leads to
\[
\de \F = \iint \dfun{\F}{\zeta(x,y)} \de \etac(x,z) \dif x
\mathrm{d}y,
\]
where $y$ is now related to $z,\etac(x,z)$ through (\ref{eqn:atri}). 
Comparing with the second equality of (\ref{eqn:afun}) gives the result
\[
\dfun{\F}{\zeta(x,y)} = \dfun{\F}{\etac(x,z)},
\]
analogous  to that of Zakharov and Piterbarg.
The bracket (\ref{eqn:breuler}) can now be expressed in terms of the
new variables:
\begin{eqnarray*}
\{\F,\G\} &=& - \iint \dfun{\F}{\zeta} \left[\zeta_0(y) + \ep
\zeta(x,y), \dfun{\G}{\zeta}\right] \dif x \mathrm{d} y \\
&=& - \iint  \dfun{\F}{\etac} \, \frac{\partial(\zeta_0(z),{\displaystyle
\dfun{\G}{\etac}})}{\partial(x,y)} \,
\left|\frac{\partial(x,y)}{\partial(x,z)}\right| \dif x \mathrm{d}z \\
&=& \iint \left.\dt{\zeta_0}{y_0}\right|_z \dfun{\F}{\etac}
\dpar{}{x}\left(\dfun{\G}{\etac}\right) \, \dif x \mathrm{d} z.
\end{eqnarray*}
Correspondingly, the Poisson  operator is given in terms of the
variables $x,z,\etac$ by $\J_0=\zetaop \partial_x$.

It is easy to see that the transformation
(\ref{eqn:atr})--(\ref{eqn:atri}) is equivalent to the
Zakharov--Piterbarg transformation when $\zeta_0(y_0)=\beta y_0$;
indeed, one finds,
\[
z = y + \ep \beta^{-1} \zeta(x,y), \quad \etac(x,z)=\zeta(x,y),
\]
which is equivalent to (\ref{eqn:zp}) and to their equation (6) (noting
the correspondence 
$(\zeta,\etac) \rightarrow (\Omega,\zeta)$ between our notation and
theirs). That our transformation (\ref{eqn:CV4}) is the leading order
approximation to (\ref{eqn:atr})--(\ref{eqn:atri}) can be verified by
expanding the expression for $z$ in (\ref{eqn:atr}) to obtain
\[
z = y + \ep \left. \dt{y_0}{\zeta_0} \right|_{\zeta_0(y)} \zeta(x,y) +
\frac{\ep^2}{2} \left. \dt{^2 y_0}{\zeta_0^2} \right|_{\zeta_0(y)}
\zeta^2(x,y) + O(\ep^3).
\]
Using the $O(\ep)$ approximation to eliminate the implicit $y$-dependence in
terms of order $O(\ep)$, one can inverse this relation according to
\[
y = z - \ep \left. \dt{y_0}{\zeta_0} \right|_{\zeta_0(z)} \zeta(x,z) +
\frac{\ep^2}{2} \left. \dt{^2 y_0}{\zeta_0^2}\right|_{\zeta_0(z)}
\zeta^2(x,z) + \frac{\ep^2}{2}
\left(\left. \dt{y_0}{\zeta_0}\right|_{\zeta_0(z)}\right)^2
\dpar{}{z}[\zeta^2(x,z)] + O(\ep^3).
\]
Introducing this result into the definition (\ref{eqn:atr}) for
$\etac$ yields
\[
\etac(x,z) = \zeta(x,z) - \frac{\ep}{2}
\dpar{}{z}\left[\left. \dt{y_0}{\zeta_0} \right|_{\zeta_0(z)}
\zeta^2(x,z) \right] + O(\ep^2),
\]
which is equivalent to (\ref{eqn:CV4}), $z$ replacing $y$.

\section{Free energy for the stratified fluid} \label{app:strat}

The free energy for stratified fluids is obtained by combining (\ref{eqn:sfham}) with 
the Casimir functionals for the bracket (\ref{eqn:kibr}) which have the form
\begin{equation}  \label{eqn:sfcas}
\C_1 = \iint C_1(\rho_0 + \ep \rho) \dif x \mathrm{d} y, \quad 
\C_2 = \iint (\zeta_0 + \ep \zeta)  C_2(\rho_0 + \ep \rho) \dif x
\mathrm{d} y,
\end{equation}
for arbitrary functions $C_1(\cdot)$ and $C_2(\cdot)$. The proper
choice can be found to be
\begin{equation} \label{eqn:sfcas2}
C_1(\cdot) = - \int_{\rho_0}^{(\cdot)} h(\mu) \dif \mu, \quad
C_2(\cdot) = \psi_0(\cdot),
\end{equation}
where $\psi_0(\cdot)$ and $h:=\zeta_0
\mathrm{d}\psi_0 \, /\mathrm{d}\rho_0 + g y - U^2/2$ are defined by their
functional dependence on $\rho_0$. This dependence follows from the
steady state condition for the basic solution.   
Adding these Casimir functionals to the Hamiltonian leads to the free
energy in the form 
\begin{eqnarray*}
\F_{(\zeta,\rho)} &=& \iint \left\{
\frac{1}{2} (\rho_0+\ep \rho) |\nabla \psi|^2 + \dt{\psi_0}{\rho_0}
\zeta \rho - \ep^{-1} \int_0^\rho [h(\rho_0+\ep \mu) - h(\rho_0)]
\dif \mu \nonumber \right. \\
&& + \left. \ep^{-2} (\zeta_0 + \ep \zeta) \left[ \psi_0(\rho_0+\ep \rho) -
\psi_0(\rho_0) - \ep \dt{\psi_0}{\rho_0} \rho \right] \right\} \dif x
\mathrm{d}y,
\end{eqnarray*}
which can be expanded in $\ep$. Keeping only the $O(1)$ and $O(\ep)$
terms, we find
\begin{eqnarray} \label{eqn:sffree}
\F_{(\zeta,\rho)} &=& \iint \left\{ \frac{1}{2} \rho_0 |\nabla \psi|^2 +
\dt{\psi_0}{\rho_0} \zeta \rho + \frac{1}{2} \left( \zeta_0 \dt{^2
\psi_0}{\rho_0^2} - \dt{h}{\rho_0} \right) \rho^2 \right. \nonumber
\\
&& + \left. \ep \left[ \frac{1}{2} \rho |\nabla \psi|^2 + \frac{1}{6}
\dt{^2\psi_0}{\rho_0^2} \zeta \rho^2 +  \frac{1}{6} \left( \zeta_0 \dt{^3
\psi_0}{\rho_0^3} - \dt{^2h}{\rho_0^2} \right) \rho^3 \right]
\right\}
\dif x \mathrm{d}y + O(\ep^2).
\end{eqnarray}

The next step is to introduce the variable transformation
(\ref{eqn:kiCV}) into this free energy. We decompose the streamfunction according to $\psi=\mphi+\ep \mtheta$ and find $\mphi$ and $\mtheta$ given by (\ref{eqn:sfphi}).
The terms in $|\nabla \psi |^2$ in (\ref{eqn:sffree}) are then readily
expressed as functions of $\mphi$ and $\mtheta$.  Substitution of
(\ref{eqn:kiCV}) into (\ref{eqn:sffree}) yields terms in $\eta \si$, $\si^2$, $\eta \si^2$ and
$\si^3$. Lengthy calculations, using extensively the relation
\[
\zeta_0 \dt{\psi_0}{\rho_0} - h = \frac{U^2}{2} - g y
\]
and its successive derivatives  are necessary to put them in a simple form.
This leads to the free energy (\ref{eqn:sffreefinal}). 

The functional derivatives of  $\F$ need to be calculated to derive the equations of motion. Using
(\ref{eqn:sfphi}), one can find
\begin{eqnarray*}
\dfun{\F_{(\eta,\si)}}{\eta} &=& - \mphi - \frac{U^\prime}{\rhoop} \si
+ \ep \left(
\frac{\mphi_y \si}{\rhoop} + \frac{U^\prime}{2 {\rhoop}^2} \si^2 -
\mtheta \right) \nonumber \\
\dfun{\F_{(\eta,\si)}}{\si} &=& - \frac{U}{\rhoop} \eta + U \mphi_y +
\frac{1}{\rhoop} \left(U U^\prime + \frac{\zetaop U}{\rhoop} - g
\right) \si \nonumber \\
&& + \ep \left\{ \frac{1}{\rhoop} \mphi_y \eta - \frac{1}{2} |\nabla
\mphi|^2 - \frac{\zetaop}{{\rhoop}^2} \mphi_y \si - \frac{1}{\rhoop} 
(U \mphi_y)_y \si \right. \nonumber \\
&& \left. + U \mtheta_y + \frac{U^\prime}{{\rhoop}^2} \eta \si -
\frac{1}{2 {\rhoop}^3} \left[ 2 \zetaop U^\prime + \rhoop (U
U^\prime)^\prime \right] \si^2 \right\}
\end{eqnarray*}

\section{Free energy for two-dimensional MHD} \label{app:mhd}

We derive the free energy for two-dimensional MHD by adding to the Hamiltonian (\ref{eqn:mhdham}) a  
suitable Casimir functionals whose general form 
is again (\ref{eqn:sfcas}). The choice leading to a quadratic free
energy is formally given by (\ref{eqn:sfcas2}), with $h = \zeta_0 \,
\mathrm{d} \psi_0 / \mathrm{d} \rho_0 - \nabla^2 \rho_0$. The general
form of the free energy is found to be 
\begin{eqnarray*}
\F_{(\zeta,\rho)} &=& \iint \left\{ 
\frac{1}{2} \left(|\nabla \psi|^2 + |\nabla \rho|^2 \right)
- \ep^{-1} \int_0^\rho \left[ h(\rho_0+\ep \mu)-h(\rho_0) \right] \dif
\mu \right. \\
&& + \left.  \ep^{-2} (\zeta_0 + \ep \zeta) \left[\psi_0(\rho_0+\ep
\rho)-\psi_0(\rho_0)-\ep \dt{\psi_0}{\rho_0} \rho \right] +
\dt{\psi_0}{\rho_0} \zeta \rho \right\} \dif x \mathrm{d}y,
\end{eqnarray*}
and after expanding in $\ep$
\begin{eqnarray} \label{eqn:mhdfree}
\F_{(\zeta,\rho)} &=& \iint \left\{ 
\frac{1}{2} \left(|\nabla \psi|^2 + |\nabla \rho|^2 \right)
+ \dt{\psi_0}{\rho_0} \zeta \rho
+ \frac{1}{2} \left(\zeta_0 \dt{^2 \psi_0}{\rho_0^2} - \dt{h}{\rho_0}
\right) \rho^2  \right. \nonumber \\
&& + \left. \ep \left[\frac{1}{2} \dt{^2\psi_0}{\rho_0^2} \zeta \rho^2 +
\frac{1}{6} \left(\zeta_0 \dt{^3 \psi_0}{\rho_0^3} - \dt{^2h}{\rho_0^2}
\right) \rho^3 \right] \right\} \dif x \mathrm{d}y + O(\ep^2).
\end{eqnarray} 
Introducing the variable change (\ref{eqn:kiCVinv}), with $\psi=\mphi+\ep \mtheta$, finally leads to the free energy in the form (\ref{eqn:mhdfree2}). 

The variational derivatives of this free energy with respect to $\eta$ and $\si$ can then
be evaluated; they are given by
\begin{eqnarray*}
\dfun{F_{(\eta,\si)}}{\eta} &=&  - \mphi - \frac{U}{\rhoop} \si + \ep \left[ -
\mtheta + \frac{1}{\rhoop} \mphi_y \si + \frac{U^\prime}{2 {\rhoop}^2}
\si^2 \right] \\
\dfun{F_{(\eta,\si)}}{\si} &=& - \nabla^2 \si - \frac{U}{\rhoop} \eta +
\frac{1}{{\rhoop}^2} \left(\zetaop U + \rhoop \rho^{(3)}\right) \si \\
&& + \ep \left[ \frac{1}{\rhoop}\mphi_y \eta -
\frac{\zetaop}{{\rhoop}^2} \phi_y \si + \frac{U^\prime}{\rhoop} \eta
\si \right. \\
&& + \left. \frac{1}{\rhoop} \si \nabla^2 \si_y - \nabla^2 \dpar{}{y}
\frac{\si^2}{2 \rhoop} - \frac{1}{{2\rhoop}^3} \left(2 \zetaop U^\prime
+ \rhoop \rho_0^{(4)} \right) \si^2 \right] \,.
\end{eqnarray*}

\section{Canonization for continuous spectra}
\label{sec:canon}

Once the weakly nonlinear equations are beatified, formulated as a Hamiltonian system
with a constant Poisson operator $\J_0$, it is natural to seek the next step of canonization, i.e., transforming  into a canonical form. Standard
(canonical) perturbation methods, starting with the diagonalization  of
the leading-order Hamiltonian, can then be applied to study (and
possibly integrate) the weakly nonlinear dynamics. Because the bracket $\J_0$
found for the weakly nonlinear system is also that of the linearized
system, both the canonization and the diagonalization of the
leading-order Hamiltonian are achieved if one uses the action--angle
variables of the linearized system. For finite-dimensional 
systems, those variables are easier to find (see Appendix \ref{app:twospin} for an example). For infinite-dimensional
systems, however, this can be a non-trivial task unless the equilibrium
solution is simple. Indeed, the spectrum of the linear
evolution operator generally contains a continuous part which needs to be
treated carefully. Nevertheless, existing results provide
action--angle variables corresponding to the continuous spectrum for
the linearized vorticity and  Vlasov--Poisson equations \cite{MP92,M00,BM02,v96,v98} that we have considered in Section \ref{sec:eul}.  Here, we show how these results can be exploited to cast the weakly
nonlinear equations in a very simple form, well-suited to start
further investigations. We focus on the Vlasov--Poisson equation, as it
is somewhat simpler than the vorticity equations. The latter could be
treated equivalently with minor modifications. 

The transformations rendering the linear Vlasov--Poisson equations into 
canonical form was  first described in  \cite{MP92} and are intimately related to the Van Kampen generalized eigenmodes of the  linearized system.    We apply these 
transformations to the variable $\eta$. Schematically, the following
successive changes of variables are made:
\[
\eta(x,v,t) \rightarrow \eta_k(v,t) \rightarrow E_k(u,t) \rightarrow
J_k(u,t),\theta_k(u,t).
\]
Here, $\eta_k$ is the Fourier transform of $\eta$ in $x$, $E_k$ are
coordinates in which the linearized system is diagnonalized, and $(J_k,\theta_k)$ are the
action--angle variables. The transformation from $\eta_k$ to $E_k$
can be seen as an integral transform defined by
\begin{equation} \label{eqn:catrd}
\eta_k(v,t) = \frac{\ic k}{4 \pi e} \Gd[E_k(u,t)] := \frac{\ic k}{4 \pi e}
\int_{-\infty}^\infty G_k(u,v) E_k(u,t) \dif u,
\end{equation}
with the kernel
\[
G_k(u,v) := \epsi(k,v) \Pr \frac{1}{u-v} + \epsr(k,v) \de(u-v),
\]
where $\epsi$ and $\epsr$ are the imaginary and real parts of the
plasma dielectric function. The inverse transform can be shown to be
\[
E_k(u,t) = \frac{4 \pi e}{\ic k} \Gi[\eta_k(v,t)] :=  \frac{4 \pi
e}{\ic k}
\int_{-\infty}^\infty \tilde{G}_k(u,v) \eta_k(v,t) \dif v,
\]
with the kernel
\[
\tilde{G}_k(u,v) := \frac{1}{|\eps(k,u)|^2}\left[ \epsi(k,u) \Pr \frac{1}{u-v} + \epsr(k,u) \de(u-v)
\right].
\]
The action/angle variables are then defined by
\[
E_k(u,t) = \sqrt{\frac{16 |\epsi|}{|k| V |\eps|^2} J_k(u,t)} \exp\left[-\ic
s_k \theta_k(u,t)\right],
\]
where $s_k := \text{sign}\,k$. In terms of these variables, the Poisson
bracket (\ref{eqn:vlJ}) becomes canonical, i.e.
\[
\{\F,\G\} = \sum_{k=1}^{\infty}  \int_{-\infty}^\infty \left(
\dfun{\F}{\theta_k} \dfun{\G}{J_k} - \dfun{\F}{J_k}
\dfun{\G}{\theta_k}
\right) \dif u,
\]
and the quadratic part of the Hamiltonian is simply
\[
\F^{(2)} = \sum_{k=1}^{\infty}  \int_{-\infty}^\infty \omega_k(u) J_k(u,t)
\dif u,
\]
with the frequencies $\omega_k(u) := k u \, \text{sign}\, \epsi(k,u)$.

To derive a canonical form of the Hamiltonian equations
(\ref{eqn:vleq}) describing the weakly nonlinear evolution of a
disturbance \cite{Y01}, we only need to express the $O(\ep)$ (cubic) part of the
Hamiltonian (\ref{eqn:vlham3}) in terms of the action--angle
variables. Substituting (\ref{eqn:catrd}) and integrating in $x$, one
finds
\begin{eqnarray} \label{eqn:calong}
&-& \frac{m}{3} \iint \frac{\eta^3}{{f_0^\prime}^2} \dif x \text{d}v
= \frac{\ic m}{3(4 \pi e)^3} \sum_{k_a} \sum_{k_b} \int \frac{k_a k_b
k_c}{{f_0^\prime}^2} \dif v \\  
&\times& \left( \int _{-\infty}^\infty G_{k_b}(u_a,v)
E_{k_a}(u_a,t) \dif u_a \int _{-\infty}^\infty G_{k_b}(u_b,v)
E_{k_b}(u_b,t) \dif u_b \int _{-\infty}^\infty G_{k_c}(u_c,v)
E_{k_c}(u_c,t) \dif u_c \right) \nonumber ,
\end{eqnarray}
with $k_a+k_b+k_c = 0$. This expression can be written in the standard
form
\[
\F^{(3)} = \sum_{k_a} \sum_{k_b} \iiint I(k_a,k_b,k_c,u_a,u_b,u_c)
E_{k_a}(u_a,t) 
E_{k_b}(u_b,t) E_{k_a}(u_c,t) \dif u_a \, \text{d} u_b \, \text{d}
u_c,
\]
where $I(\cdots)$ is an interaction coefficient. To derive the
expression for
this coefficient from (\ref{eqn:calong}), one needs to permute the
integrations with respect to $v$ and $u_a,u_b,u_c$. Because of the
presence of the Cauchy principal value in $G_k(u,v)$, the
Poincar\'e--Bertrand equality must be used. After a lenghty
calculation, the details of which can be found in \cite{Y01}, one finds
\[
I =  \frac{\ic m k_a k_b k_c}{3(4 \pi e)^3} \int_{-\infty}^{\infty} j(u_a,u_b,u_c,v)
\dif v,
\]
with
\begin{eqnarray*}
j &=& \frac{1}{{f_0^\prime}^2}  G_{k_a}(u_a,v)  G_{k_b}(u_b,v)  
G_{k_c}(u_c,v) \\ 
&+& \frac{1}{{f_0^\prime}^2} \left[ \epsi(k_a,v) \epsi(k_b,v) \de(v-u_a)
\de(v-u_b) G_{k_c}(u_c,v) + \text{ cyc. } \right],
\end{eqnarray*}
where cyc designates the sum of the three cyclic permutations of
$(a,b,c)$.


\bibliographystyle{unsrt}

\bibliography{Hpeturb1}

\end{document}